\documentclass[aps,prb,twocolumn,showpacs,amsmath,amssymb,superscriptaddress,floatfix]{revtex4-2}

\usepackage[english]{babel}
\usepackage{blindtext}
\usepackage{latexsym}
\usepackage{amssymb}
\usepackage{physics}
\usepackage{amsmath}
\usepackage{bm}
\usepackage{amsfonts}
\usepackage{relsize}
\usepackage{xcolor}
\usepackage{verbatim}
\usepackage{bbold}
\usepackage{slashed}
\usepackage{appendix}
\usepackage{graphicx}
\usepackage[normalem]{ulem}
\usepackage[colorlinks = true,
            linkcolor = blue,
            urlcolor  = blue,
            citecolor = blue,
            anchorcolor = blue]{hyperref}
\usepackage{graphicx}
\usepackage{dcolumn} 
\newcommand{\approptoinn}[2]{\mathrel{\vcenter{
  \offinterlineskip\halign{\hfil$##$\cr
    #1\propto\cr\noalign{\kern2pt}#1\sim\cr\noalign{\kern-2pt}}}}}

\newcommand{\bs}[1]{\mathbf{#1}}

\newcommand{\bq}{\bs{q}}
\newcommand{\bQ}{\bs{Q}}

\newcommand{\eps}{\epsilon}

\definecolor{ao(english)}{rgb}{0.0, 0.5, 0.0}
\definecolor{amaranth}{rgb}{0.9, 0.17, 0.31}
\definecolor{green(html/cssgreen)}{rgb}{0.0, 0.5, 0.0}

\newcommand\greensout{\bgroup\markoverwith{\textcolor{green(html/cssgreen)}{\rule[0.5ex]{2pt}{1.0pt}}}\ULon}

\begin{document}
\title{Comprehensive mean-field analysis of magnetic and charge orders in the two-dimensional Hubbard model}
\author{Robin Scholle}
\affiliation{Max Planck Institute for Solid State Research, D-70569 Stuttgart, Germany}
\author{Pietro M.~Bonetti}
\affiliation{Max Planck Institute for Solid State Research, D-70569 Stuttgart, Germany}
\author{Demetrio Vilardi}
\affiliation{Max Planck Institute for Solid State Research, D-70569 Stuttgart, Germany}
\author{Walter Metzner}
\affiliation{Max Planck Institute for Solid State Research, D-70569 Stuttgart, Germany}
\date{\today}
\begin{abstract}
We present an unbiased mean-field analysis of magnetic and charge orders in the two-dimensional Hubbard model on a square lattice, both at zero and finite temperatures. Unrestricted Hartree-Fock calculations on large finite lattices are complemented by solutions restricted to N\'eel and circular spiral order in the thermodynamic limit. The magnetic states are classified by a systematic scheme based on the dominant Fourier components of the spin texture.
On finite lattices a whole zoo of ordering patterns appears. We show that many of these states are finite size artifacts related to the limited choice of ordering wave vectors on a finite lattice. In the thermodynamic limit only three classes of states with a relatively simple structure survive: N\'eel, circular spiral, and stripe states. Stripes involve also charge order and can be unidirectional or bidirectional, with horizontal and/or vertical orientation. We present complete phase diagrams in the plane spanned by electron density and temperature, for a moderate Hubbard interaction and various choices of the next-nearest neighbor hopping amplitude.
\end{abstract}
\maketitle


\section{Introduction}

The two-dimensional Hubbard model is a prototype for competing ordering tendencies in interacting electron systems. It captures key features of the valence electrons in high-$T_c$ cuprates such as antiferromagnetism and d-wave superconductivity \cite{Scalapino2012}. In spite of substantial progress in the development of quantum many body methods, only fragments of the phase diagram of this important model have been established \cite{Arovas2022,Qin2022}.

There is no doubt that at half-filling the ground state of the two-dimensional Hubbard model with pure nearest neighbor hopping is a N\'eel ordered antiferromagnet for any repulsive Hubbard interaction $U>0$, and beyond a certain critical repulsion if hopping amplitudes beyond nearest neighbors are present.
However, for electron densities below half-filling the N\'eel state quickly becomes unstable \cite{Shraiman1989, Shraiman1992, Chubukov1992, Chubukov1995}, leading to other magnetic states or possibly phase separation. These states are usually metallic and thus prone to pairing instabilities.

There is a whole zoo of possible magnetic states. Except for the N\'eel state, most of them are characterized by one or several ordering wave vectors which are generically incommensurate with the lattice periodicity. Many approximate calculations point toward planar circular spirals \cite{Shraiman1989, Shraiman1992, Chubukov1992, Chubukov1995, Dombre1990, Fresard1991, Kotov2004, Igoshev2010, Yamase2016, Eberlein2016, Mitscherling2018, Bonetti2020a} and collinear spin density waves \cite{Schulz1989, Zaanen1989, Machida1989, Poilblanc1989, Schulz1990, Kato1990, Seibold1998, Fleck2000, Fleck2001, Raczkowski2010, Timirgazin2012, Peters2014} as the most important candidates. The latter entail charge order with regions of reduced charge density arranged along one-dimensional lines, and are therefore also known as ``spin-charge stripes'' \cite{Kivelson2003}. By contrast, the charge distribution of (circular) spiral states remains uniform.

Recently, exact numerical calculations on finite lattices have provided strong evidence for stripe order in the ground state of the Hubbard model with pure nearest neighbor hopping and large $U$ for special hole-doping fractions: $1/10$ and $1/8$, corresponding to electron densities $0.9$ and $0.875$, respectively \cite{Zheng2017,Qin2020}. No superconductivity was found in these cases. However, allowing for next-nearest neighbor hopping, superconductivity reemerges for a range of densities away from half-filling \cite{Jiang2019}.
Exact numerical evaluations at finite temperatures yield peaks in the spin susceptibility at the N\'eel wave vector $(\pi,\pi)$ near half-filling, and peaks at incommensurate wave vectors in the spin and charge susceptibility for sufficiently large hole doping \cite{Simkovic2022, Xiao2023}.

Unlike superconductivity, magnetic order in the (repulsive) Hubbard model can be captured already within a static mean-field approximation, also known as Hartree-Fock theory. While the size of the order parameter and the regime of ordered states in the phase diagram are grossly overestimated by mean-field theory, finite size effects can be studied in more detail because much larger lattices can be dealt with. For N\'eel, circular spiral, and stripe states, mean-field calculations can be performed directly in the thermodynamic limit, albeit only for wave vectors commensurate with the lattice in the case of stripes.
Indeed, numerous Hartree-Fock studies of the two-dimensional Hubbard model have already appeared. However, in many of them the magnetic states were restricted to  ferromagnetic and N\'eel \cite{Lin1987, Hofstetter1998, Langmann2007}, or spiral order \cite{Igoshev2010}. The latter includes ferromagnetic and N\'eel states as the special cases with wave vectors $(0,0)$ and $(\pi,\pi)$, respectively. For small hole doping, this restriction often leads to phase separation in paramagnetic, ferromagnetic, and antiferromagnetic regions \cite{Langmann2007, Igoshev2010}.
Allowing for arbitrary collinear spin order or even for completely arbitrary spin configurations -- combined with charge order -- spin-charge stripes have been found \cite{Schulz1989, Zaanen1989, Machida1989, Poilblanc1989, Schulz1990, Kato1990, Timirgazin2012,Matsuyama2022}.

With very few exceptions, Hartree-Fock studies of the two-dimensional Hubbard model have been limited to the ground state. At finite temperatures, any magnetic order is prohibited by thermal fluctuations, as is rigorously established by the Mermin-Wagner theorem \cite{Mermin1966}. Magnetic states obtained in Hartree-Fock approximation at finite temperatures obviously violate this theorem. However, such states become meaningful as an ingredient for theories of fluctuating magnetic order, where the electron is fractionalized into a ``chargon'' with a magnetically ordered pseudospin, and a fluctuating spin rotation matrix, which restores the SU(2) spin symmetry of the electronic state \cite{Scheurer2018, Sachdev2019review, Bonetti2022gauge}.
The ``handshake'' between magnetic order found in the ground state and magnetic correlations at finite temperatures has attracted much interest recently, and there has been some progress in attempts to close the gap between exact numerical solutions obtained from zero and finite temperature algorithms \cite{Wietek2021, Simkovic2022a, Xiao2023}.

In this article we perform a comprehensive and completely unrestricted Hartree-Fock study of the two-dimensional Hubbard model at zero and finite temperatures. We choose a moderate interaction strength and analyze a broad range of electron densities below, at, and above half-filling. We consider both the ``pure'' Hubbard model, with pure nearest neighbor hopping $t$, as well as its physically more relevant extension with next-nearest neighbor hopping $t'$. Unrestricted real space calculations on large lattices (from $20 \times 20$ to $48 \times 48$) are complemented by momentum space calculations for N\'eel and spiral states in the thermodynamic limit.

As our main result we present three complete mean-field phase diagrams spanned by density and temperature, for three choices of $t'/t$. All of them contain N\'eel, spiral, and stripe states, where, for $t' \leq -0.15$, the latter two are confined to the hole doped region (below half-filling). Other, more complex magnetic states appearing in the real space calculations could be identified as finite size artifacts, which are related to the restriced discrete choice of ordering wave vectors on the finite lattice. The momentum space calculations for N\'eel and spiral states are consistent with the real space calculations, and instabilities toward stripe states are signalled by a divergence of the spin-charge susceptibility in the spiral state.

The remainder of the paper is structured as follows. In Sec.~\ref{sec: Method} we describe the real-space mean-field theory, and in Sec.~\ref{sec: Results} we present our results. In the Conclusion in Sec.~\ref{sec: Conclusion} we provide a summary and outlook. Some details on the calculations and classification of states are described in two appendices.


\section{Model and method} \label{sec: Method}

The Hubbard Hamiltonian for spin-$\frac{1}{2}$ fermions with intersite hopping amplitudes $t_{jj'}$ and a local interaction $U$ reads
\begin{equation} \label{eq: HubbardHamiltonian}
 H = H_0 + H_{\text{int}} =
 \sum_{j,j',\sigma} t_{jj'} c^\dagger_{j\sigma} c^{\phantom\dagger}_{j'\sigma} +
 U \sum_j n_{j\uparrow} n_{j\downarrow} \, ,
\end{equation}
where $c_{j\sigma}$ ($c^\dagger_{j\sigma}$) annihilates (creates) an electron on lattice site $j$ with spin orientation $\sigma \in \{\uparrow,\downarrow\}$, and $n_{j\sigma} = c^\dagger_{j\sigma} c^{\phantom\dagger}_{j\sigma}$. The hopping matrix $t_{jj'}$ depends only on the distance between the sites $j$ and $j'$, and it is parametrized by $-t$ if $j$ and $j'$ are nearest neighbor sites, and by $-t'$ if $j$ and $j'$ are next-to-nearest neighbors. We consider only the repulsive Hubbard model where $U$ is positive. In the following, we use the nearest neighbor hopping amplitude $t$ as our energy unit, that is, we set $t=1$.


\subsection{Real-space mean-field theory}

In the mean-field (or Hartree-Fock) approximation, the interaction part of the Hamiltonian $H_\mathrm{int}$ is replaced by an effective quadratic (in $c$, $c^\dagger$) operator. Following Zaanen and Gunnarsson \cite{Zaanen1989}, we decouple $H_\mathrm{int}$ as
\begin{eqnarray} \label{eq: InteractionTermHubbardModelMeanField}
 H_{\text{int}}^\mathrm{MF} &=&
 \sum_{j\sigma} \Delta_{j\overline{\sigma}} n_{j\sigma}
  + \sum_j \left( \Delta_{j-} c^{\dagger}_{j\uparrow} c_{j\downarrow}
    + \Delta_{j+} c^\dagger_{j\downarrow} c_{j\uparrow}\right) \nonumber \\
 && -\frac{1}{U}\sum_j \left( \Delta_{j\uparrow} \Delta_{j\downarrow}
   - \Delta_{j-} \Delta_{j+} \right) ,
\end{eqnarray}
with the convention $\overline{\uparrow} = \downarrow$ and $\overline{\downarrow} = \uparrow$.
The parameters $\Delta_{j\alpha}$ are self-consistently determined by the "gap equations"
\begin{subequations}\label{eq: GapVectorDefiningEquation}
    \begin{align}
        &\Delta_{j\sigma} = U\langle n_{j\sigma}\rangle,\\
        &\Delta_{j+} = \Delta_{j-}^* = -U\langle c_{j\uparrow}^\dagger c_{j\downarrow} \rangle.    
    \end{align}
\end{subequations}
The decoupling in Eq.~\eqref{eq: InteractionTermHubbardModelMeanField} allows for arbitrary spin and charge patterns. In most of the previous mean-field studies of the Hubbard model, only the first (Hartree) or the second (Fock) term on the right hand side of Eq.~\eqref{eq: InteractionTermHubbardModelMeanField} have been considered, restricting the spin order to collinear or spiral order, respectively.

We iteratively solve the mean-field equations on a finite square lattice with $\mathcal{N}_x$ ($\mathcal{N}_y$) sites in the $x(y)$-direction and periodic boundary conditions. We denote the total number of lattice sites by $\mathcal{N} = \mathcal{N}_x \mathcal{N}_y$.

The mean-field Hamiltonian can be written as
\begin{equation} \label{MeanFieldHamburger}
 H^\mathrm{MF} = H_0 + H_\mathrm{int}^\mathrm{MF} =
 \sum_{j,j'} \sum_{\sigma,\sigma'}
 c^\dagger_{j\sigma}\mathcal{H}_{jj'}^{\sigma\sigma'}c_{j'\sigma'} + \mbox{const},
\end{equation}
where $\mathcal{H}_{jj'}^{\sigma\sigma'}\in \mathbb{C}^{2\mathcal{N}\times2\mathcal{N}}$ is a square matrix with $4\mathcal{N}$ real parameters to be determined self-consistently ($\Delta_{j\uparrow}$, $\Delta_{j\downarrow}$, $\mathrm{Re}\Delta_{j+}$, and $\mathrm{Im}\Delta_{j+}$ for each lattice site). 
These parameters are related to the expectation values of the charge and spin order parameters by
\begin{subequations}
 \begin{align}
  & \langle n_j\rangle = \frac{\Delta_{j\uparrow}+\Delta_{j\downarrow}}{U}, \\
  & \langle S^z_j\rangle=\frac{\Delta_{j\uparrow}-\Delta_{j\downarrow}}{2U} , \\
  & \langle S^x_j\rangle = -\frac{\Delta_{j+}+\Delta_{j-}}{2U} , \\
  & \langle S^y_j\rangle = -\frac{\Delta_{j+}-\Delta_{j-}}{2i U} ,
 \end{align}
\end{subequations}
where $n_j = n_{j\uparrow} + n_{j\downarrow}$ and $\Vec{S}_j = \frac{1}{2} c^\dagger_j  \Vec{\sigma} c_j$, with $\vec{\sigma}$ the Pauli matrices.

We run calculations at fixed density $n$, that is, we enforce the constraint $\frac{1}{\mathcal{N}}\sum_{j\sigma} \langle n_{j\sigma} \rangle = n$. In the ground state this is achieved by calculating the expectation values on the right-hand side of Eq.~\eqref{eq: GapVectorDefiningEquation} as
\begin{equation} \label{eq: exp value T=0}
 \langle c^\dagger_{j\sigma}c_{j\sigma'}\rangle =
 \sum_{\ell=1}^N (v^\ell_{j\sigma})^* v^\ell_{j\sigma'},
\end{equation}
where $N$ is the total particle number, $\ell$ labels the $2\mathcal{N}$ eigenvalues $\eps_\ell$ of $\mathcal{H}_{jj'}^{\sigma\sigma'}$ in ascending order, and $v^\ell_{j\sigma}$ is the normalized eigenvector corresponding to the $\ell$-th eigenvalue. In this way, only rational densities of the form $n = N/\mathcal{N}$ are obtained.
At finite temperatures $T>0$, one needs to introduce a chemical potential $\mu$, and Eq.~\eqref{eq: exp value T=0} is replaced by
\begin{equation} \label{eq: exp value T!=0}
    \langle c^\dagger_{j\sigma}c_{j\sigma'}\rangle = \sum_{\ell=1}^{2\mathcal{N}} (v^\ell_{j\sigma})^* v^\ell_{j\sigma'} f(\eps_\ell-\mu),
\end{equation}
with $f(x)=1/(1+e^{x/T})$ the Fermi function. The chemical potential is adjusted to enforce the chosen value for the average density
\begin{equation} \label{eq: Find chemical potential}
    \frac{1}{\mathcal{N}}\sum_j\langle n_j\rangle=\frac{1}{\mathcal{N}}\sum_{\ell=1}^{2\mathcal{N}} f(\eps_\ell-\mu)=n,
\end{equation}
where we have made use of the property $\sum_{j,\sigma}(v_{j\sigma}^\ell)^*v_{j\sigma}^\ell=1$, 
deriving from the orthonormality condition of the eigenvectors.

We solve our system of equations in the following iterative manner:
Starting from a random initial set of parameters $\Delta_{j\alpha}$ with
$\alpha \in \{ \uparrow,\downarrow,+,- \}$ and $\Delta_{j-} = \Delta_{j+}^*$, we plug them into $\mathcal{H}_{jj'}^{\sigma\sigma'}$, from which we compute the eigenvectors and the expectation values in Eq.~\eqref{eq: GapVectorDefiningEquation} to update the values of $\Delta_{j\alpha}$. We repeat this procedure until convergence is reached. At finite temperatures, we compute the chemical potential at every iteration from Eq.~\eqref{eq: Find chemical potential}. More details on the numerical implementation can be found in Appendix~\ref{sec: ImplementationDetails}.

Within the mean-field approximation, the free energy per lattice site is given by
\begin{equation}\label{eq: free energy}
 \begin{split}
 F/\mathcal{N} = & -\frac{T}{\mathcal{N}}
 \sum_\ell \log(1 + e^{-\beta (\epsilon_\ell - \mu)}) \\
 & -\frac{1}{\mathcal{N}U}\sum_j
 \left( \Delta_{j\uparrow}\Delta_{j\downarrow} - \Delta_{j-}\Delta_{j+}\right) +
 \mu n.
 \end{split}
\end{equation}
%


\subsection{Classification of magnetic orders}\label{ClassificationSection}

\begin{figure*}
 \includegraphics[width=1.\textwidth]{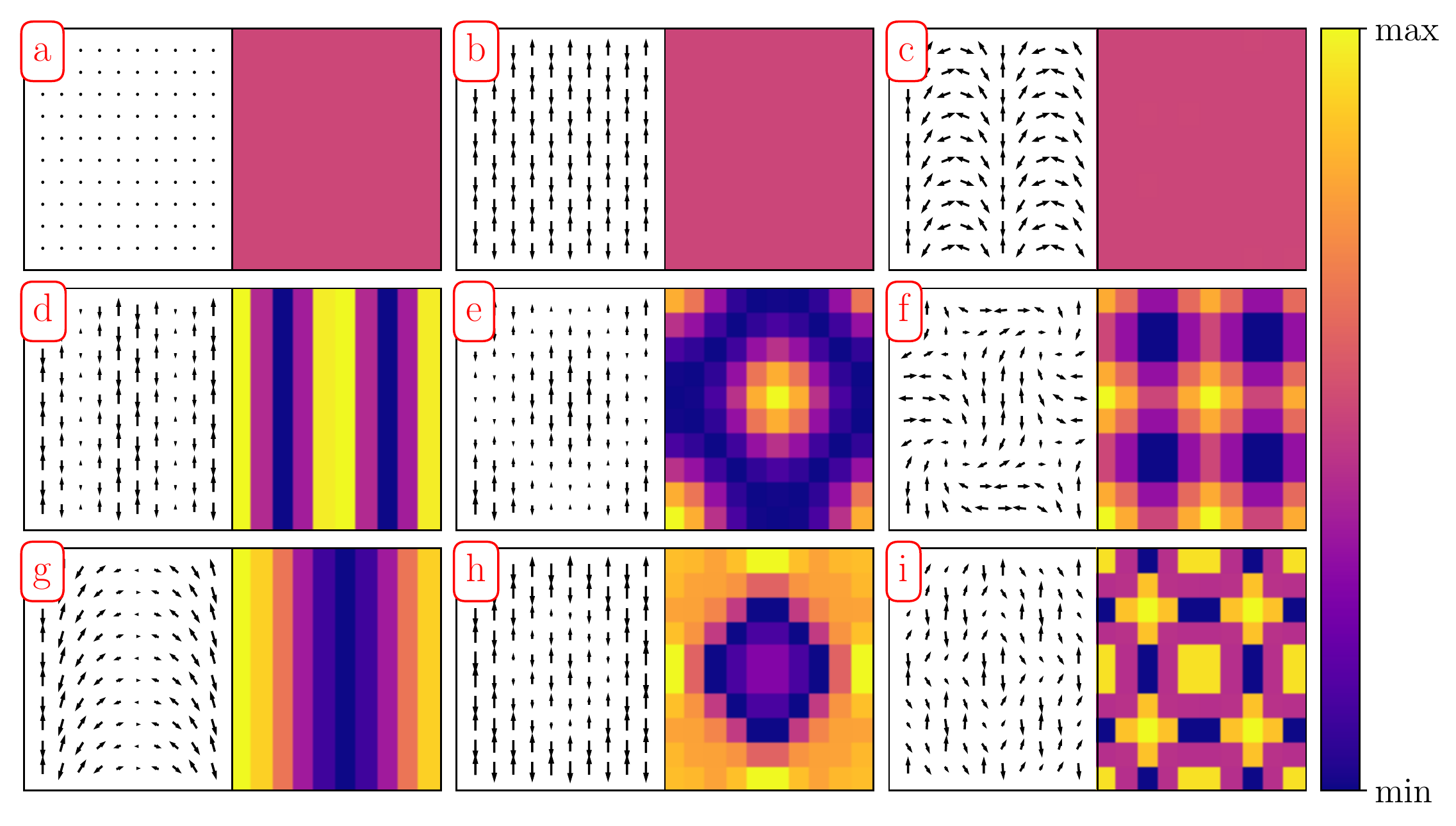}
 \caption{Overview of the different orders found in our calculations, schematically shown on a $10 \times 10$ lattice. In each panel, the square on the left shows the relative spin orientations and amplitudes (length of the arrows) of each phase, where we chose a frame such that the spins lie the $x$-$y$ plane and the bottom left spin points along the $y$-direction. The right plot in each panel shows the corresponding charge modulation, using a color code defined on the right edge of the figure. The various panels exemplify the following magnetic orders: (a) paramagnetism, (b) N\'eel antiferromagnetism, (c) spiral order, (d) stripe order, (e) collinear bidirectional stripe order, (f) coplanar bidirectional stripe order, (g) beat order, (h) other collinear orders, (i) ``strange'' order. Only one example of strange order has been shown, while we also find different ones, often with less regular patterns.}
 \label{fig: Panel_With_Different_Orders}
\end{figure*}
In this section we introduce a classification of the distinct magnetic states we find. We classify the different states solely based on their spin pattern, similarly to the classification of Sachdev \emph{et al.}~\cite{Sachdev2019}. In agreement with Ref.~\cite{Sachdev2019}, we generally find that the modulation of the charge density is roughly proportional to the square of the spin amplitudes, that is,
\begin{equation}\label{eq: charge equal spin^2}
 \langle n_j \rangle \approx
 \mbox{const} \, \big\langle \Vec{S}^2_j \big\rangle + \mbox{const} ,
\end{equation}
as could be deduced from the lowest order coupling between the charge and spin density wave order parameters in a Landau theory~\cite{ZacharKivelson1998}.

To identify a state from a (converged) real space spin pattern defined by the spin expectation values $\langle \Vec{S}_j \rangle$, we perform a Fourier transform,
\begin{equation}\label{Classification_state_ala_Sachdev}
 \big\langle \Vec{S}_j \big\rangle =
 \sum_{\bq} \Vec{\mathcal{S}}_{\bq}\, e^{i\bq\cdot \mathbf{r}_j},
\end{equation}
where $\Vec{\mathcal{S}}_{\bq}$ with $\vec{\mathcal{S}}_\bq = \vec{\mathcal{S}}^*_{-\bq}$ is the Fourier component of the spins corresponding to the wave vector $\bq$, and $\mathbf{r}_j$ represents the real space coordinate of lattice site $j$. The sum runs only over the $\mathcal{N}=\mathcal{N}_x \mathcal{N}_y$ momenta allowed by the periodic boundary conditions on the finite size lattice. Since we are exclusively dealing with even numbers $\mathcal{N}_x$ and $\mathcal{N}_y$, the allowed momenta can be written as
\begin{equation}\label{Introduce_pitch}
 \bq \in \left\{ \left. \left(
 \pi - \pi\frac{2\nu_x}{\mathcal{N}_x},\pi-\pi\frac{2\nu_y}{\mathcal{N}_y} \right)
 \right|
 \nu_\alpha \in \left\{ 0,1,\dots,\mathcal{N}_\alpha-1\right\} \right\} .
\end{equation}
In the majority of cases only one or two modes with fixed wave vectors $\bq$ (together with their partner $-\bq$) contribute significantly to $\langle \vec{S}_j \rangle$.
Usually, these wave vectors have the form $\bQ_x \equiv (\pi-2\pi\eta_x,\pi)$ and/or $\bQ_y \equiv (\pi,\pi-2\pi\eta_y)$. The parameters $\eta_\alpha$ are frequently refered to as ``incommensurabilities'' in the literature.
Note that, in a finite system, $\eta_\alpha$ is restricted to integer multiples of $1/\mathcal{N}_\alpha$ (see Eq.~\eqref{Introduce_pitch}).
For lattices with $\mathcal{N}_x = \mathcal{N}_y$, we always find $\eta_x = \eta_y \equiv \eta$.
In cases with only one mode, we can express $\vec{\mathcal{S}}_\bq$ as
\begin{equation} \label{eq: one mode}
 \vec{\mathcal{S}}_\bq = \frac{1}{2} 
 \left( \vec{\mathcal{S}} \, \delta_{\bq,\bQ} +
 \vec{\mathcal{S}}^* \, \delta_{\bq,-\bQ} \right) , 
\end{equation}
where $\bQ$ has the form $\bQ_x$ or $\bQ_y$. In states with two modes, we have
\begin{eqnarray} \label{eq: two modes}
 \vec{\mathcal{S}}_\bq &=& 
 \frac{1}{2} \left( \vec{\mathcal{S}}_x \, \delta_{\bq,\bQ_x} +
 \vec{\mathcal{S}}_x^* \, \delta_{\bq,-\bQ_x} \right) \nonumber \\ &+&
 \frac{1}{2} \left( \vec{\mathcal{S}}_y \, \delta_{\bq,\bQ_y} +
 \vec{\mathcal{S}}_y^* \, \delta_{\bq,-\bQ_y} \right) .
\end{eqnarray}

In our mean-field calculations we find 9 distinct phases. We sketch the spin and charge order pattern for a representative of each phase in Fig.~\ref{fig: Panel_With_Different_Orders}. Phases that can be transformed into each other by a point group symmetry operation of the lattice (rotations and reflections), by a spatial translation (changing the phases of $\vec{\mathcal{S}}_x$ and $\vec{\mathcal{S}}_y$), or by a global SU(2) rotation of the spin frame, belong to the same class. In the following, we describe each class by one of its representatives.

In principle, there might be other possible spin configurations, for example the phases $F$ and $G$ in the classification of Ref.~\cite{Sachdev2019}, but we do not find them in the parameter regime chosen in our calculations.


\subsubsection{Paramagnetism}

In the paramagnetic phase the expectation value of the spin on each site vanishes, that is,
$\langle \Vec{S}_j \rangle = 0$, and the charge distribution is uniform, $\langle n_j \rangle = n$ (see panel~(a) of Fig.~\ref{fig: Panel_With_Different_Orders}).

\subsubsection{N\'eel antiferromagnetism}

In a N\'eel antiferromagnet only $\Vec{\mathcal{S}}_{(\pi,\pi)}$ is nonzero. In real space, adjacent spins point in opposite directions, all spins have the same amplitude, and the charge is homogeneously distributed, as shown in panel~(b) of Fig.~\ref{fig: Panel_With_Different_Orders}.

\subsubsection{Spiral}

We refer to planar circular spiral states briefly as ``spiral'' states.
Such a state is characterized by a single mode with a wave vector $\bQ$, and (as a representative) 
\begin{equation}
 \Vec{\mathcal{S}} = \mathcal{S}_0 \begin{pmatrix} 1 \\ i \\ 0 \end{pmatrix} ,
\end{equation}
where $\mathcal{S}_0$ is a real number.
In real space, the corresponding spin pattern has the form 
\begin{equation}
 \big\langle \Vec{S}_j \big\rangle = \mathcal{S}_0
 \begin{pmatrix}
 \cos(\bQ \cdot \mathbf{r}_j) \\
 \sin(\bQ \cdot\mathbf{r}_j) \\
 0
 \end{pmatrix}.
\end{equation}
The spin magnitude $|\langle \Vec{S}_j \rangle|$ is constant (independent of $j$), and the charge distribution is therefore homogeneous, as shown in panel~(c) of Fig.~\ref{fig: Panel_With_Different_Orders}. The wave vector $\bQ$ has the form $\bQ_x = (\pi-2\pi\eta,\pi)$ or $\bQ_y = (\pi,\pi-2\pi\eta)$, that is, only one component deviates from $\pi$.
The N\'eel state can be viewed as the special case of a spiral with $\bQ = (\pi,\pi)$.

\subsubsection{Unidirectional stripes}

A (unidirectional) stripe state is described by a single $\bq$-mode such that 
\begin{equation}
 \vec{\mathcal{S}} = \mathcal{S}_0 \begin{pmatrix} 1 \\ 0 \\ 0 \end{pmatrix}.
\end{equation}
In this state, the spins are ordered collinearly with a modulation along the $x$- (or $y$-) axis. In real space, the spin modulation is given by 
\begin{equation}\label{eq: PERFECT STRIPES}
 \big\langle \Vec{S}_j \big\rangle = \mathcal{S}_0
 \begin{pmatrix}
 \cos(\bQ \cdot \mathbf{r}_j) \\ 0 \\ 0 \end{pmatrix} .
\end{equation}
Once again, the wave vector $\bQ$ has either the form $\bQ_x$ or $\bQ_y$.
The charge order of this state is a unidirectional charge density wave with a wave vector $2\bQ$, that is, $\langle n_j \rangle - n \propto \cos(2\bQ \cdot \mathbf{r}_j)$. The minima in the amplitude of the spin coincide with the minima of the charge modulation, as shown in panel~(d) of Fig.~\ref{fig: Panel_With_Different_Orders}.

\subsubsection{Collinear bidirectional stripes (ClBS)}
We call a state a collinear bidirectional stripe if it consists of two orthogonal stripes with parallel spin orientations, that is, if the nonzero components of $\vec{\mathcal{S}}_\bq$ in Eq.~\eqref{eq: two modes} are given by
\begin{equation}
 \Vec{\mathcal{S}}_{x} = \Vec{\mathcal{S}}_{y} =
 \mathcal{S}_0 \begin{pmatrix} 1 \\ 0 \\ 0 \end{pmatrix} ,
\end{equation}
where $\mathcal{S}_0$ is a real constant.
In real space, this corresponds to
\begin{equation}
 \big\langle \Vec{S}_j \big\rangle =
 \mathcal{S}_0 \begin{pmatrix}
 \cos(\bQ_x\cdot \mathbf{r}_j)+\cos(\bQ_y\cdot \mathbf{r}_j) \\
 0 \\ 0 \end{pmatrix}.
\end{equation}
This phase has charge modulations in both $x$- and $y$-directions, proptional to $\left[ \cos(\bQ_x \cdot \mathbf{r}_j) + \cos(\bQ_y\cdot \mathbf{r}_j) \right]^2$, as displayed in panel~(e) of Fig.~\ref{fig: Panel_With_Different_Orders}.

\subsubsection{Coplanar bidirectional stripes (CpBS)}

We call a state a coplanar bidirectional stripe if it consists of two orthogonal stripes with orthogonal spin orientations, that is, if we find two $\bq$-modes of the form 
\begin{equation}
 \Vec{\mathcal{S}}_{x} = 
 \mathcal{S}_0 \begin{pmatrix} 1 \\ 0 \\ 0 \end{pmatrix}, \quad
 \Vec{\mathcal{S}}_{y} = 
 \mathcal{S}_0 \begin{pmatrix} 0 \\ 1 \\ 0 \end{pmatrix} ,
\end{equation}
with $\mathcal{S}_0$ a real constant. 
In real space representation this corresponds to
\begin{equation}
    \big\langle \Vec{S}_j \big\rangle = \mathcal{S}_0
    \begin{pmatrix}
    \cos(\bQ_x \cdot \mathbf{r}_j) \\
    \cos(\bQ_y \cdot \mathbf{r}_j)\\
    0
    \end{pmatrix}.
\end{equation}
This phase also has charge modulations in both $x$- and $y$-directions, proportional to 
$[\cos(\bQ_x \cdot \mathbf{r}_j)]^2 + [\cos(\bQ_y \cdot \mathbf{r}_j)]^2$, as shown in panel~(f) of Fig~\ref{fig: Panel_With_Different_Orders}.

\subsubsection{Beat states}

Sometimes we find states whose dominant $\bq$-components take the form (up to a lattice rotation) $\bQ_x=(\pi-2\pi\eta,\pi)$ and $\bQ'_x=(\pi-2\pi\eta',\pi)$ with $\eta\neq \eta'$.
As shown in panel~(g) of Fig.~\ref{fig: Panel_With_Different_Orders}, in these states there are unidirectional charge modulations and the spin patterns usually show beat-like modulations, since the spin components are sums of cosines with different wave numbers and phases.
As we will discuss in more detail in Sec.~\ref{sec: EnergyConsiderations}, we attribute these states to finite size effects. In fact, they arise due to the fact that the "optimal" incommensurability $\eta_\mathrm{opt}$ lies in between $\eta$ and $\eta'$, which is however not allowed by the finite size of the system (cf.~Eq.~\eqref{Introduce_pitch}). In the thermodynamic limit, where any (real) value of $\eta$ is permitted, we expect these states to converge to a pure stripe or spiral phase. 

\subsubsection{Other collinear orders}

In some cases all the spins are ordered collinearly (that is, $\langle \vec{S}_j\rangle \propto \hat{n}$ with $\hat{n}$ a constant unit vector), but not in the shape of N\'eel order, stripe order or collinear bidirectional stripes.
In these phases, the amplitude of the spins varies spatially and as a result we also obtain a charge modulation. 
As noted in Ref.~\cite{Zaanen1989}, these states may be replaced by stripe states in the thermodynamic limit. Indeed, the lattice sizes treated in our calculations are too small to resolve stripes with wave vectors $\bQ$ close to $(\pi,\pi)$, so that the system prefers to order in a more complex manner. Such a state, sketched in panel (h) of Fig.~\ref{fig: Panel_With_Different_Orders}, can be viewed as a kind of stripe order where the lines of constant density, instead of lying parallel to the $x$- (or $y$-) direction, close onto themselves, forming a ring.
In these cases, we find several nonzero components of $\Vec{\mathcal{S}}_{\bq}$.


\subsubsection{Strange orders}

In some cases we cannot classify a state by the previously described orders.
Such states do not exhibit a simple structure in momentum space and often show complex charge patterns. In the following we will refer to these orders as ``strange order''. In the majority of cases, but not always, their spin order is coplanar. An example is sketched in panel (i) of Fig.~\ref{fig: Panel_With_Different_Orders}. We expect these strange states to disappear in the thermodynamic limit, as we will discuss more exhaustively in Sec.~\ref{sec: EnergyConsiderations}.


\section{Results} \label{sec: Results}

We now present our results. Unrestricted real-space Hartree-Fock calculations have been carried out on a $20 \times 20$ lattice, supplemented by a few calculations on larger lattices (up to $48 \times 48$) to analyze finite size effects. We chose $\mathcal{N}_x = \mathcal{N}_y$ to not bias the system towards orders breaking point group symmetries.
We chose a moderate coupling strength $U = 3t$ in all calculations, and three distinct ratios $t'/t$, namely $0$, $-0.15$, and $-0.3$.
We performed a temperature and filling scan with steps of 0.01 in $T/t$ and $n$.
We classified the converged states according to the classification scheme described in Sec.~\ref{ClassificationSection}. In Appendix~\ref{app: class of states}, we specify the criteria and numerical method employed to achieve this.

In most cases the system converges to a unique (modulo symmetries) state independently of the initial conditions. To reduce the risk of converging to a metastable state corresponding to a local but not global minimum of the free energy, we repeated the calculations for each point $(n,T)$ starting five times from a random configuration and then from the converged solutions at the neighboring points $(n\pm0.01,T)$ and $(n,T\pm0.01t)$. In the rare cases where the iterations converged to inequivalent states, we retained the state with the lowest free energy.

For N\'eel and, more generally, spiral states, the mean-field equations can be solved in the thermodynamic limit, using a momentum space representation \cite{Chubukov1992, Chubukov1995, Dombre1990, Fresard1991, Igoshev2010}, and the stability of the spiral states in the thermodynamic limit can be probed by computing the spin and charge susceptibilities \cite{Kampf1996, Bonetti2022}. We exploit this complementary approach to benchmark and to interprete the results from the real space finite size calculations.


\subsection{Results for $t^\prime = -0.15t$} \label{sec: t_prime_15}

We start by presenting exemplary results for a next-nearest neighbor hopping $t' = -0.15t$ at intermediate interaction strength $U = 3t$. In Sec.~\ref{sec: The other tprimes}, we show results for $t' = 0$ and $t' = -0.3t$.


\subsubsection{Phase diagram}\label{PhaseDiagramSection}

\begin{figure}[t]
 \centering
 \includegraphics[width=0.5\textwidth]{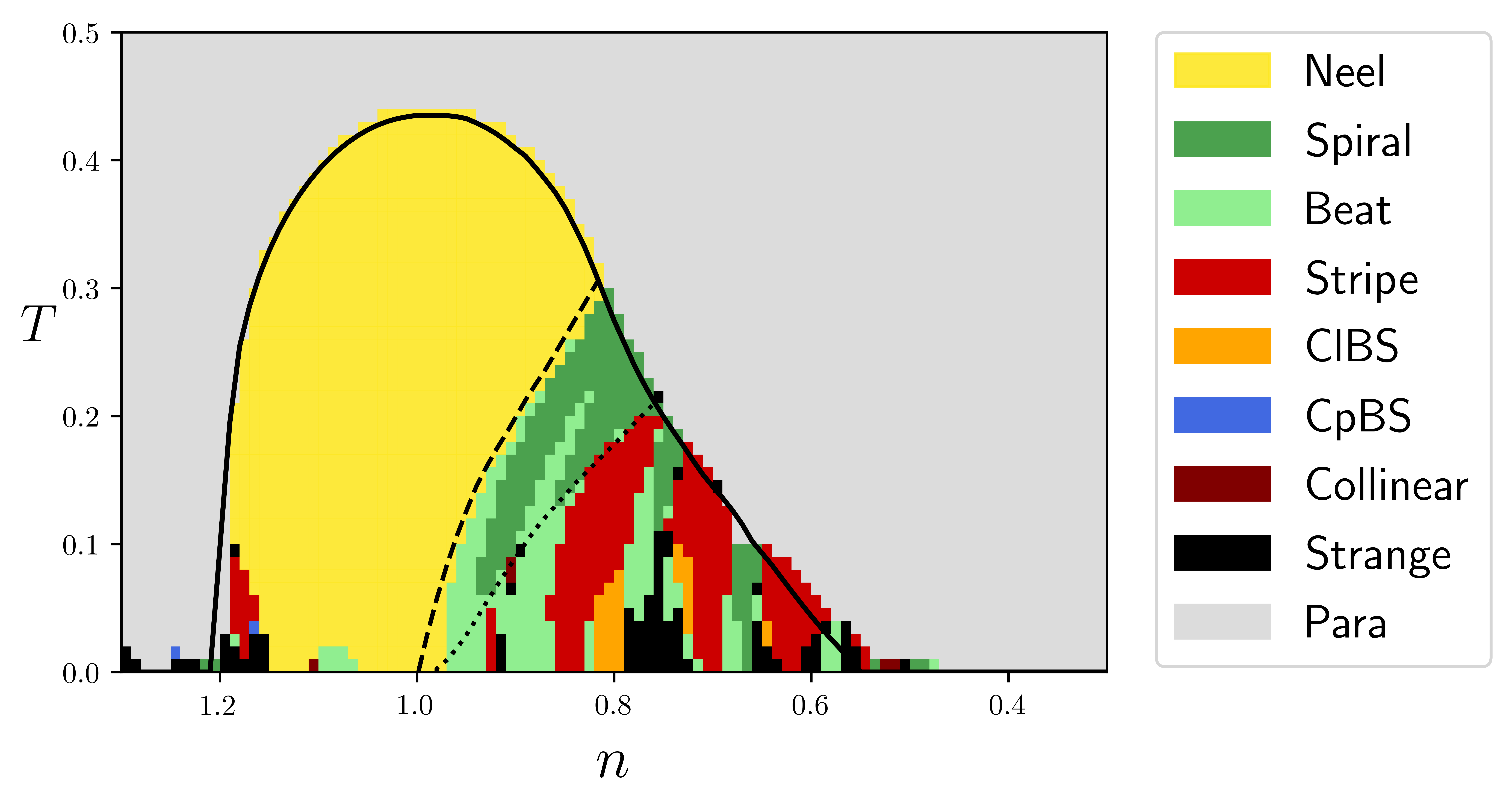}
 \caption{Phase diagram for $U=3t$ and $t' = -0.15t$. The colors label the states resulting from the real space calculation on a $20 \times 20$ lattice. The black lines were obtained from calculations in momentum space in the thermodynamic limit.
 The solid black line indicates the transition temperature $T^*$ separating the paramagnetic from the magnetically ordered regime, the dashed black line the transition between N\'eel and non-N\'eel spiral order, and the dotted black line the divergence of the charge susceptibility in the spiral state.}
\label{PhaseDiagram_tprime15}
\end{figure}
In Fig.~\ref{PhaseDiagram_tprime15}, we show the mean-field phase diagram as a function of the particle density $n$ and the temperature $T$. The colors label the states resulting from the real space calculation on a $20 \times 20$ lattice, while the black lines were obtained from calculations in momentum space in the thermodynamic limit. The transition temperature $T^*$ obtained from the momentum space calculation (solid black line) is very close to the transition temperature obtained for the finite system.
Since electronic correlations are neglected, the Hartree-Fock approximation overestimates the size of the critical dopings at the edges of the ordered regime. For the same parameters, significantly smaller critical values are found in a renormalized mean-field theory with effective interactions obtained from a functional renormalization group flow \cite{Yamase2016}.

For densities close to half-filling, we find a N\'eel-ordered region, which at $n=1$ extends up to a critical temperature $T^* \approx 0.43t$. With few exceptions, N\'eel order is also obtained in the entire electron-doped region (for $n>1$). Not only $T^*$, but also the boundary of the N\'eel ordered region inside the magnetic regime obtained from the momentum space calculation (dashed line) agrees very well with the real-space finite size calculation.

On the hole-doped side, the momentum space calculation in the thermodynamic limit yields a regime of stable spiral order with a continuously varying incommensurability $\eta$ between the dashed and the dotted black lines in Fig.~\ref{PhaseDiagram_tprime15}. In the real space calculation on the $20 \times 20$ lattice, the spiral regime is split into spiral and beat order regions, where the incommensurabilities in the spiral regions are either $\eta = 0.05$ (close to the N\'eel region) or $\eta = 0.1$.
 
At densities and temperatures where we observe beat states it is energetically more costly for the system to order as a spiral state with the discrete incommensurabilities 0.05 or 0.1 allowed on the $20 \times 20$ lattice. We will discuss this point further in Sec.~\ref{sec: EnergyConsiderations}. Since intermediate ordering vectors are prohibited due to the periodic boundary conditions, the system orders in a beat ordered pattern. Looking at the wavevectors of the Fourier modes $\Vec{\mathcal{S}}_{\bq}$ in these states, one finds $\bQ_1 = (\pi-2\pi\frac{1}{20},\pi)$ and $\bQ_2 = (\pi-2\pi\frac{1}{10},\pi)$.
We therefore expect the beat states to be artifacts of the finite system size. We recalculated the magnetic and charge orders in the beat region between the spiral states on a $40 \times 40$ lattice and found that beat order is indeed replaced by spiral order with an intermediate wavevector $\bQ = (\pi-2\pi\frac{3}{40},\pi)$. Hence, the real space calculation is consistent with a spiral phase with a smoothly varying $\eta$ (increasing upon decreasing density and temperature) in the thermodynamic limit, as suggested by the momentum space analysis.

Below the dotted line in Fig.~\ref{PhaseDiagram_tprime15}, and for any temperature at densities below $n \approx 0.75$, spiral states exhibit negative (for some wave vectors) spin and charge susceptibilities, and are thus unstable. In this region we find extended phases with stripes and other collinear orders. The stripe region is split into several domains with distinct incommensurabilities: $\eta = 0.05$ for densities $n \approx 0.93$, $\eta = 0.1$ for densities close to $n \approx 0.82$, $\eta = 0.15$ for densities around $0.7$, and $\eta = 0.2$ near $n \approx 0.62$. These stripe regimes are separated by intermediate regions of beat ordered and strange phases, where the predominant wavevectors are the ones of the two adjacent stripe regimes. In these intermediate regions, we recalculated the states on a $40 \times 40$ lattice for a few $(n,T)$ pixels and found stripes with intermediate values of $\eta$.
This confirms the hypothesis that beat and strange states are just artifacts of the finite system size.

\begin{figure}[t]
\centering
 \includegraphics[width=0.5\textwidth]{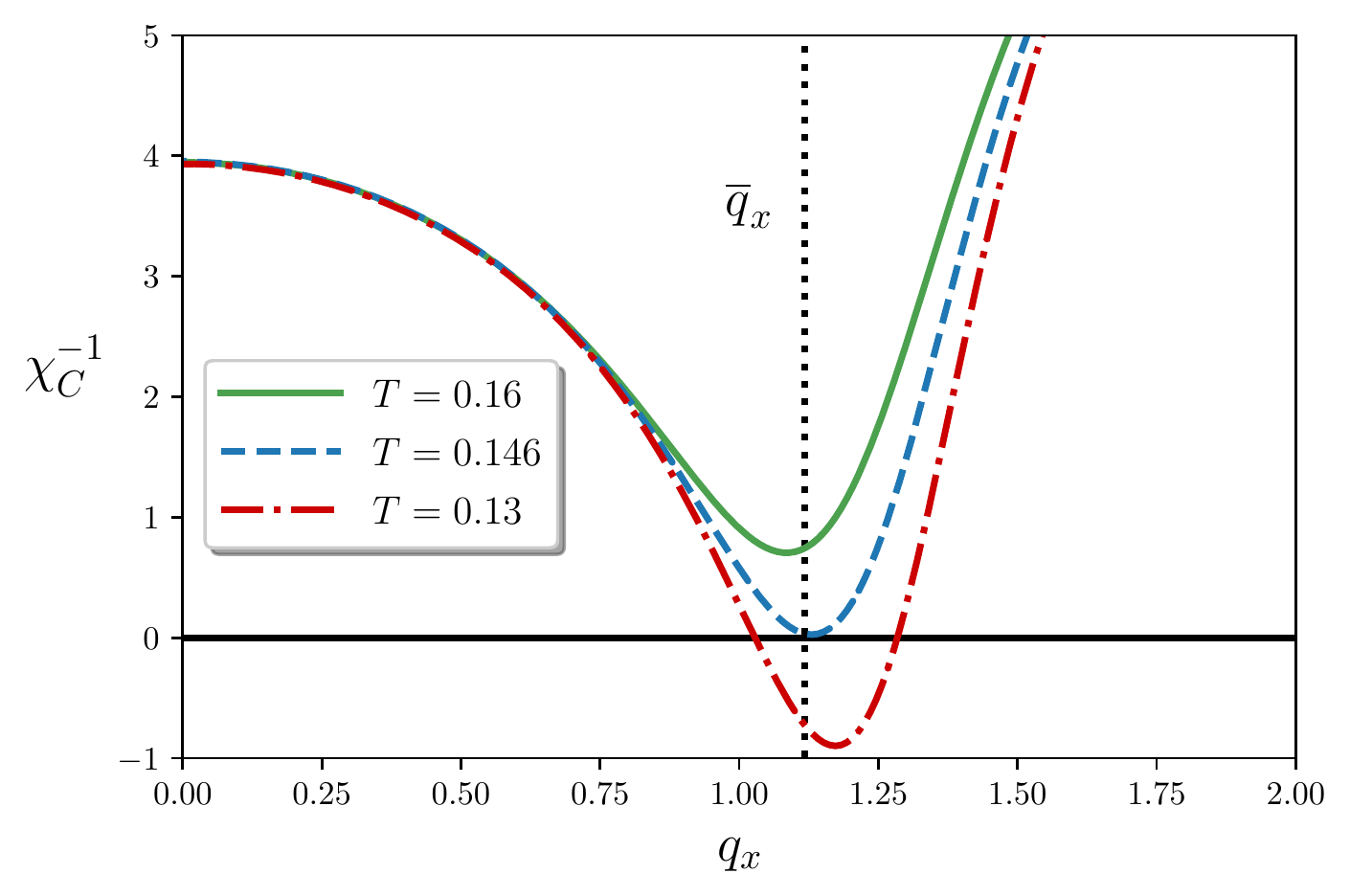}
 \caption{Inverse of the RPA charge susceptibility $\chi_C^{-1}$ along the momentum line $\bq = (q_x,0)$ for $n = 0.85$, $t' = -0.15t$ and $U = 3t$. The divergence of $\chi_C$ occurs at $T \approx 0.146$. For these parameters, we find an optimal wave vector $\bQ$ of the form $(\pi-2\pi\eta,\pi)$ with $\eta = 0.089$. We show $\chi_C^{-1}$ for three different temperatures. $\chi_C$ becomes negative for $T<0.146$.}
\label{fig: Charge_Sus}
\end{figure}

The stability of the spiral state can be probed in the thermodynamic limit by computing the spin and charge susceptibilities in momentum space in random phase approximation (RPA), following Refs.~\cite{Kampf1996,Bonetti2022}. Joint divergences are found for the in-plane spin susceptibility and for the charge susceptibility.
The dotted black line in Fig.~\ref{PhaseDiagram_tprime15} marks the line at which the charge susceptibility $\chi_C$ diverges. For temperatures below that line, one obtains an unphysical negative charge susceptibility. The divergence of the charge susceptibility, occurring for a wavevector of the form $\bQ_C = (\pm\bar{q}_x,0)$ for a spiral with wave vector $(\pi-2\pi\eta,\pi)$, is always concomitant with a divergence of the spin susceptibility in the plane in which the spirals lie. This indicates the instability of the spiral state towards a state displaying a charge modulation and a rearrangement of the spins in the plane. The fact that the line of diverging charge susceptibility coincides with the transition from spiral order to stripe order confirms the idea that stripes emerge as an instability of the spiral states~\cite{Shraiman1989,Dombre1990,Chubukov_comment}. We show $\chi_C$ for the density $n = 0.85$ and temperatures in the vicinity of the spiral-stripe transition in Fig.~\ref{fig: Charge_Sus}. Close to the transition, we find that $\bQ_C$ is given by twice the optimal spiral wavevector $\bQ$ modulo a reciprocal lattice vector, indicating the onset of stripe order with a charge modulation wavevector $2\bQ$, and a spin modulation with wavevector $\bQ$, in agreement with Eq.~\eqref{eq: charge equal spin^2}.

For densities below $n \approx 0.75$ we observe a {\em direct} second-order transition from the paramagnetic state to a stripe state. This occurs without a diverging charge susceptibility at the critical temperature (only the magnetic susceptibility diverges), as the charge modulation in the stripe phase is a second order effect, induced by the spin ordering~\cite{ZacharKivelson1998}. In the momentum space mean-field calculation with spiral order, we observe in this regime diverging charge and in-plane spin susceptibilities at any temperature below $T^*$, that is, as soon as the spiral order parameter forms. This shows that spiral order is not even metastable in the stripe regime.


\subsubsection{Magnetization}

\begin{figure}[h]
\centering
 \includegraphics[width=0.5\textwidth]{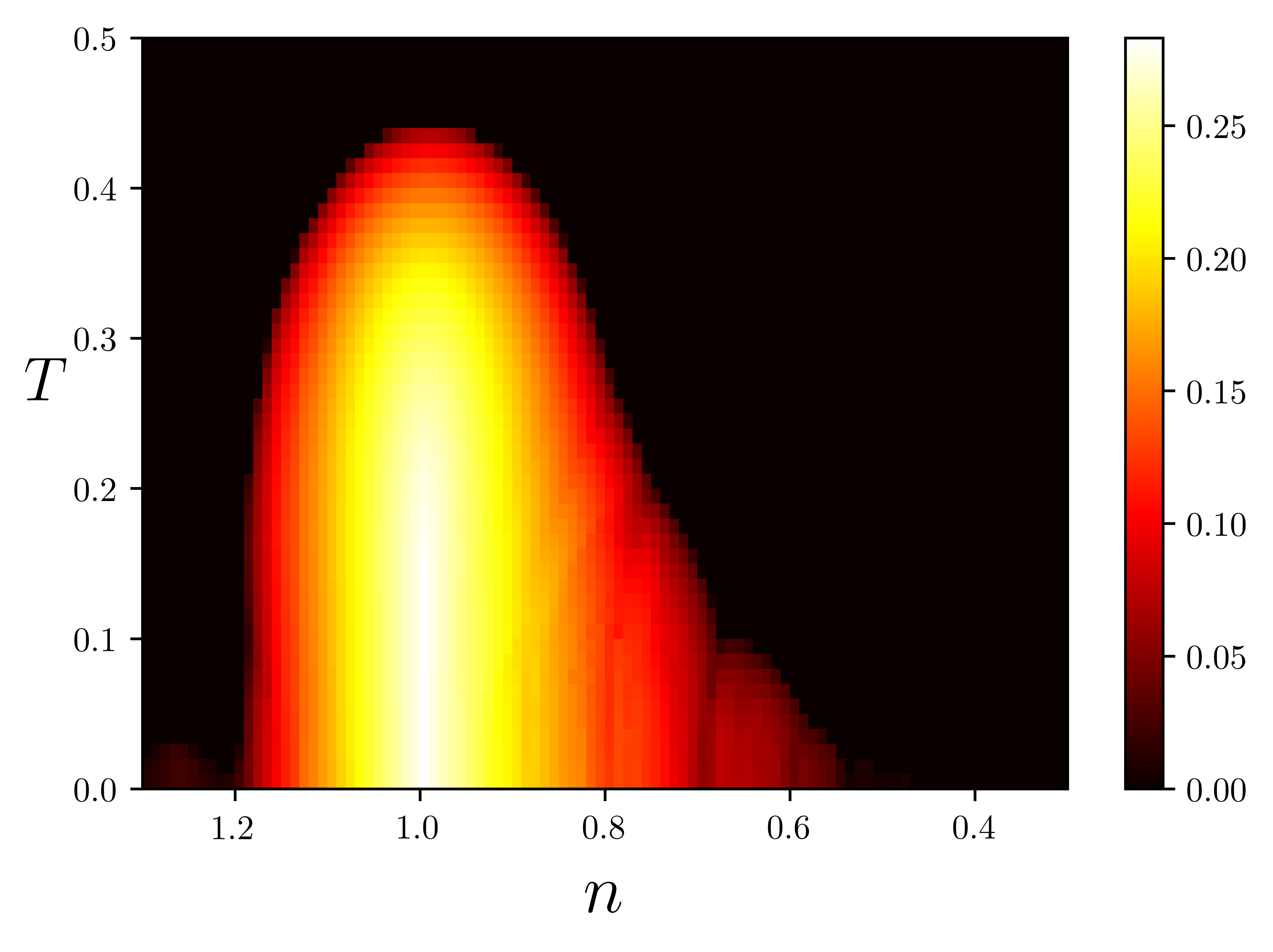}
 \caption{Average magnetization $m$ for $U=3t$ and $t^\prime = -0.15t$, calculated on a $20\times 20$ lattice.}
\label{SpinDiagram0}
\end{figure}
In Fig.~\ref{SpinDiagram0}, we plot the average magnetization per site,
\begin{equation}
    m = \frac{1}{\mathcal{N}} \sum_j \sqrt{\langle S^x_{j}\rangle^2+\langle S^y_{j}\rangle^2+\langle S^z_{j}\rangle^2}.
\end{equation}
Within the limitations due to the discrete grid for $n$ and $T$, the transitions from the paramagnetic state to N\'eel, spiral, and stripe order look all continuous.
A continuous phase transition from a paramagnetic state to a collinear spin configuration is consistent with a Landau theory of phase transitions Ref.~\cite{ZacharKivelson1998}.
The magnetization is maximal near half filling, as expected. We see slight dips in the average magnetization near the transition from stripe to spiral phases as well as in some of the beat phases. We believe that this originates from the finite size of the system, since the scale of the dips is approximately the same as the change in the value of the magnetization between calculations on a $40 \times 40$ and a $20 \times 20$ lattice.


\subsubsection{Energetic analysis}
\label{sec: EnergyConsiderations}

\begin{figure}[h]
\centering
 \includegraphics[width=0.5\textwidth]{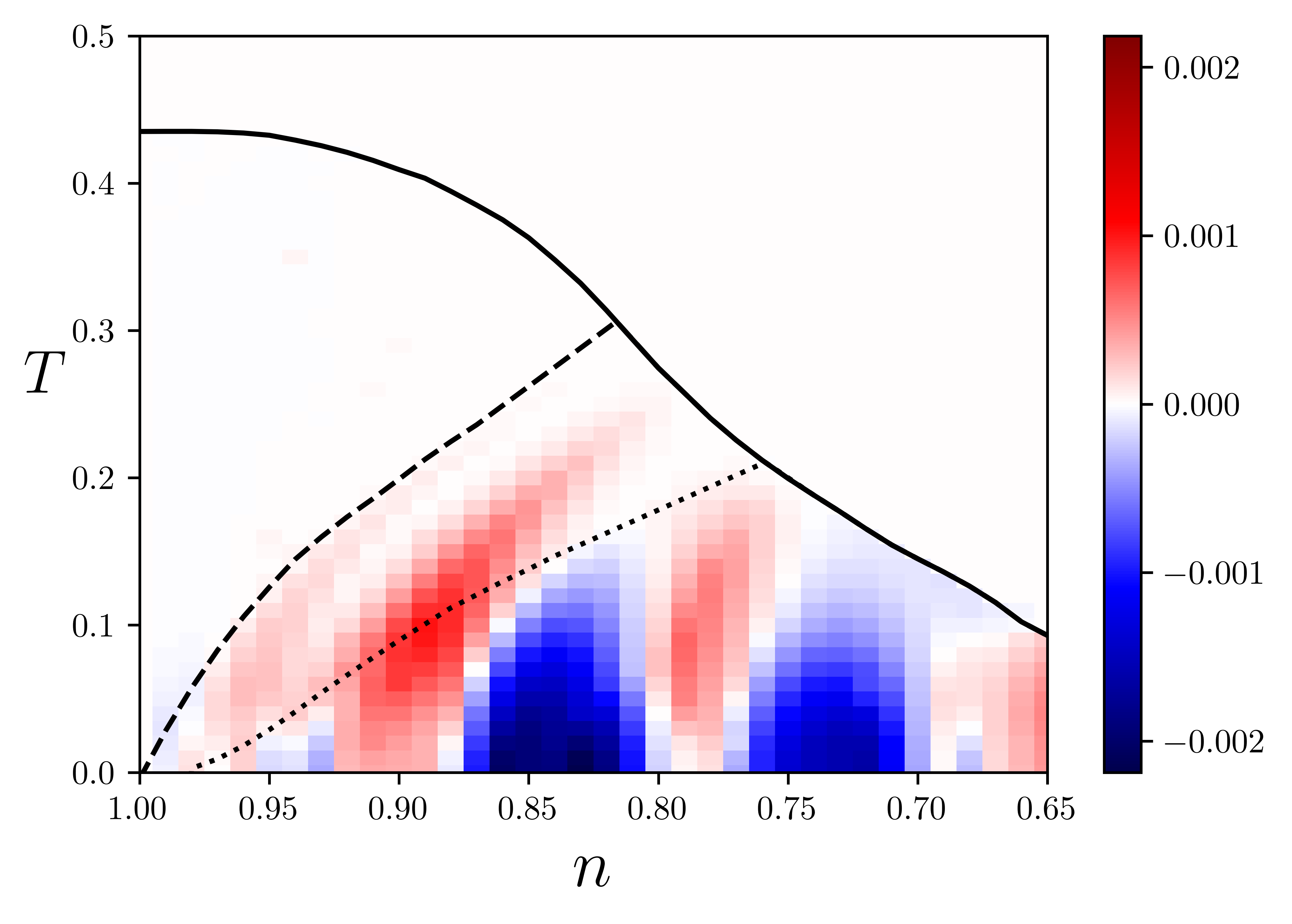} 
 \caption{Energy difference $F_\mathrm{fs} - F_\mathrm{is}$ between the real space calculation on a $20 \times 20$ lattice and the momentum space calculation in the thermodynamic limit. The model parameters are $U = 3t$ and $t^\prime =-0.15$.
 The solid black line indicates $T^*$, the dashed black line the transition between N\'eel and spiral order and the dotted black line the divergence of the charge susceptibility in the spiral state as obtained from the momentum space mean-field code (as in Fig.~\ref{PhaseDiagram_tprime15}).}
\label{EnergyDiagram015}
\end{figure}

To further clarify which phases found in our finite-size calculation will survive in the thermodynamic limit, we compared the finite-size free energy $F_\mathrm{fs}$ from Eq.~\eqref{eq: free energy} with its counterpart $F_\mathrm{is}$ in an infinite system, where the magnetic states are restricted to N\'eel and spiral order. Focusing on the more interesting hole-doped regime, in Fig.~\ref{EnergyDiagram015} we show $F_\mathrm{fs} -F_\mathrm{is}$ for densities $0.65 \leq n \leq 1$. We only show the energy differences in the magnetic regime and set the energy differences in the paramagnetic phase to zero.

To set a reference scale, we note that the typical ``condensation'' energy $F_\mathrm{fs}-F_\mathrm{pm}$, where $F_\mathrm{pm}$ is the free energy obtained in the (unstable) paramagnetic mean-field solution, is of the order of $-0.05t$. The energy difference $F_\mathrm{fs} - F_\mathrm{is}$ is of the order of a few percent of the total condensation energy. In regions where the real space calculation yields N\'eel or spiral order, $|F_\mathrm{fs} - F_\mathrm{is}|$ is smaller than $10^{-5}t$.

In the beat state separating N\'eel and spiral order, one can see that $F_\mathrm{fs} > F_\mathrm{is}$, so the beat order is not expected to be stable in the thermodynamic limit, since the spiral state with optimal $\eta$ has a lower free energy.
The same applies to the beat states separating the two spiral regions with different pitches.

Below the dotted line in Fig.~\ref{EnergyDiagram015}, we see a clear gain in energy in the centres of the stripe ordered phases, indicating that stripe order can remain stable even in the thermodynamic limit. In the intermediate beat- and strange-ordered states, where stripe orders with $\bQ$ near the optimal wavevector are prohibited by the periodic boundary conditions, we again find $F_\mathrm{fs} > F_\mathrm{is}$. Therefore, the beat- and strange-ordered states will not be stable in the thermodynamic limit. As we know from the RPA susceptibilities that also spiral states are unstable in this regime, we expect stripe order with optimized intermediate wavevectors to take their place.

The analysis of the wavevectors of the beat states described in Sec.~\ref{PhaseDiagramSection} together with the free energy analysis above make us believe that the beat states and the strange states are artifacts of the finite system size, and that in the thermodynamic limit the phase diagram consists only of a N\'eel ordered dome around half filling accompanied by a spiral region and a stripe-ordered phase with smoothly varying wave vectors $\bQ$ on the hole-doped side.


\subsection{Results for $t^\prime = 0$ and $-0.3t$} \label{sec: The other tprimes}

In this section, we show results similar to those of Sec.~\ref{sec: t_prime_15} for next-to-nearest neighbor hopping strengths $t' = 0$ and $t' = -0.3t$. We choose the same interaction strength $U = 3t$ and lattice size $\mathcal{N}_x = \mathcal{N}_y = 20$ as before.


\subsubsection{Phase diagrams}

\begin{figure}[ht]
\centering
 \includegraphics[width=0.5\textwidth]{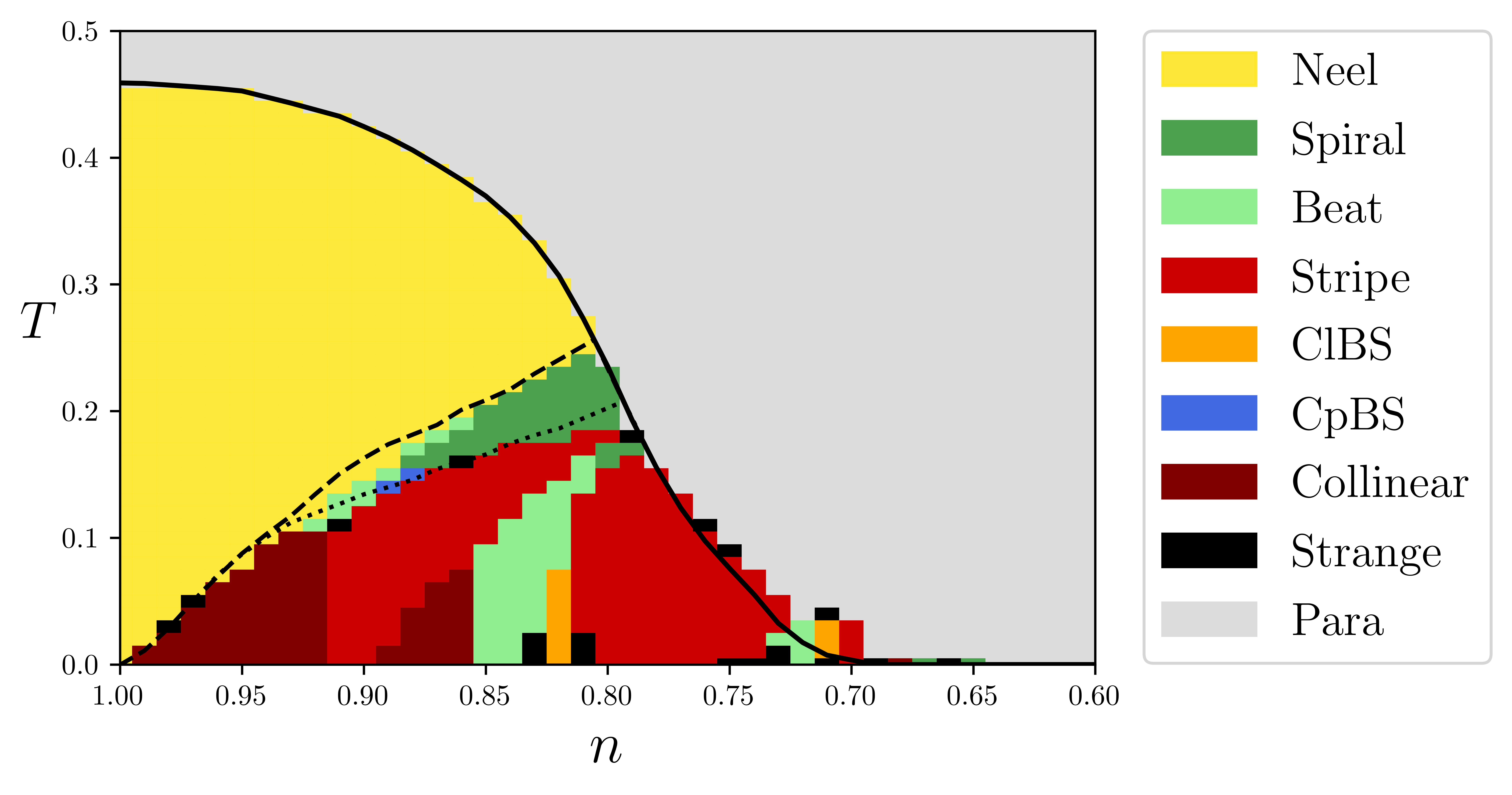}
 \caption{Phase diagram for $U=3t$ and $t' = 0$. The colors label the states obtained from a real space calculation on a $20\times 20$ lattice, and the black lines are defined as in Fig.~\ref{PhaseDiagram_tprime15}. We show the phase diagram only for densities $n \leq 1$, since the system is particle-hole symmetric without next-to-nearest neighbor hopping.}
\label{PhaseDiagram_tprime0}
\end{figure}
\begin{figure}[h]
\centering
 \includegraphics[width=0.5\textwidth]{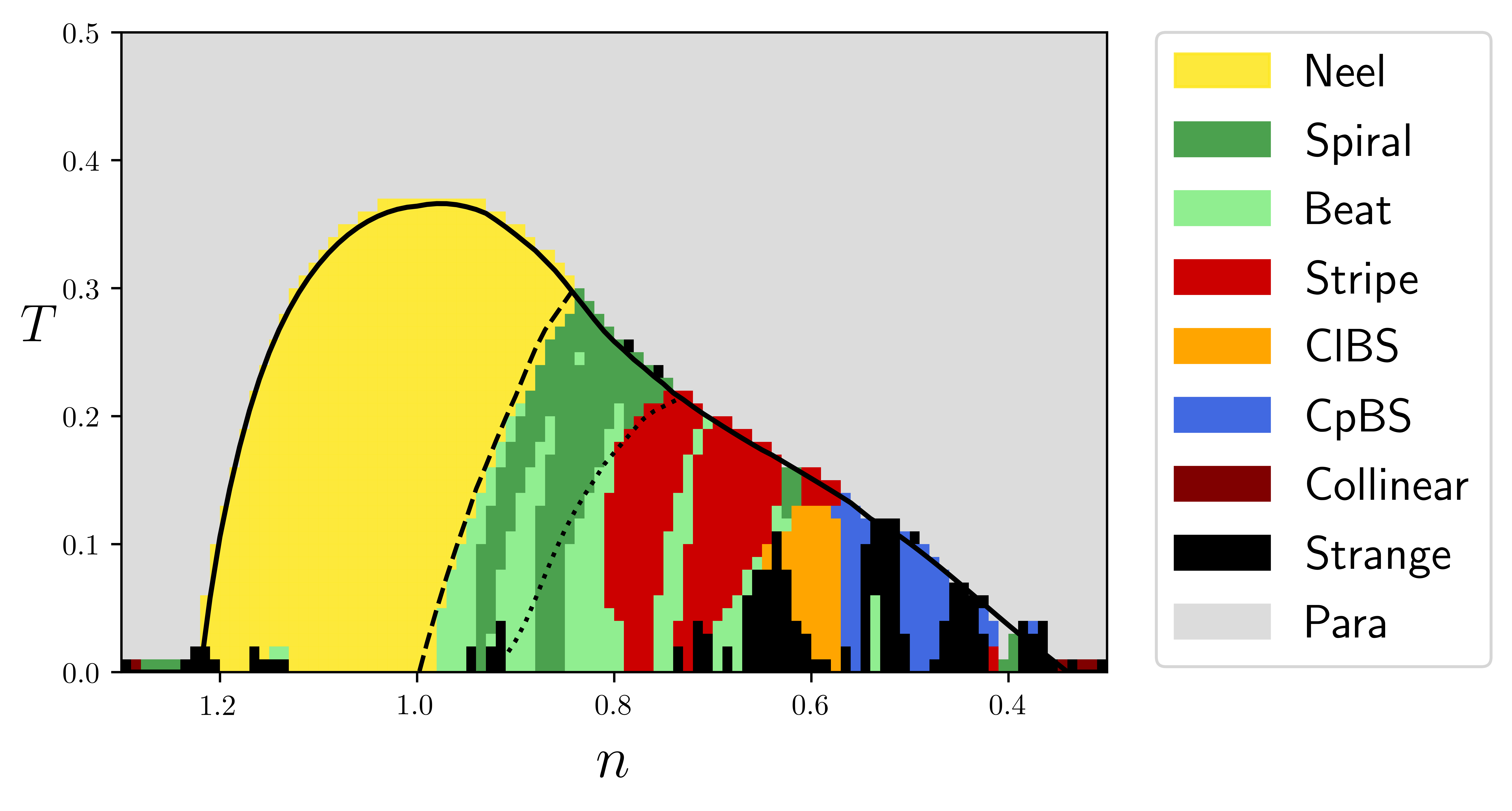}
 \caption{Phase diagram for $U=3t$ and $t' = -0.3t$. The colors label the states obtained from a real space calculation on a $20\times 20$ lattice, and the black lines are defined as in Fig.~\ref{PhaseDiagram_tprime15}.}
\label{PhaseDiagram_tprime30}
\end{figure}
The phase diagrams for $t' = 0$ and $t' = -0.3t$ are shown in Figs.~\ref{PhaseDiagram_tprime0} and \ref{PhaseDiagram_tprime30}, respectively. For $t' = 0$, the Hamiltonian is particle-hole symmetric, so that the hole- and electron-doped sides of the phase diagram look identical. Hence, only the hole-doped side of the phase diagram is shown in Fig.~\ref{PhaseDiagram_tprime0}.

In the electron doped regime, the phase diagram for $t' = -0.3t$ exhibits primarily a N\'eel ordered regime, similar to the case $t' = -0.15t$, and in agreement with previous works \cite{Igoshev2010,Yamase2016,Bonetti2022}.

On the hole doped side the critical hole doping increases for larger $|t'|$, while $T^*$ at half-filling decreases.
For all three values of $t'$, we find a N\'eel ordered regime also in the hole-doped regime in the proximity of half-filling, shrinking however at low temperatures. For lower densities, all three phase diagrams exhibit an intermediate spiral regime. Analogously to Fig.~\ref{PhaseDiagram_tprime15}, we also show the N\'eel-spiral transition line (dashed line) and the divergence of the charge susceptibility (dotted line) in the thermodynamic limit, as obtained from the momentum space calculation.
Similarly to the case $t' = -0.15t$, we find that the spiral states obtained on the $20 \times 20$ lattice are mostly located between these two lines. For $t' = 0$, all spiral states have an incommensurability of $\eta = 0.05$, whereas for $t' = -0.3t$, the spiral regime is again split into two regions, one with $\eta = 0.05$ and one with $\eta = 0.1$, separated by an intermediate beat region, in which the dominant wavevectors correspond to $\eta = 0.05$ and $0.1$.
The bigger $|t'|$ is, the larger is the area where spirals are the energetically favored state. While for $t' = 0$ the spiral area is a relatively thin strip below N\'eel order, for $t' = -0.3t$ the spiral region extends over a wide range of densities and spiral order remains stable even in the ground state for densities down to $n = 0.91$.
In the thermodynamic limit we expect a continuous evolution of $\eta$ from $\eta = 0$ to $\eta \approx 0.1$ within the spiral region, and we expect the beat states to disappear from the phase diagram.

Below the dotted line, the phase diagrams exhibit stripe ordered patches divided by intermediate beat states. For $t^\prime = 0$, there is a $\eta = 0.05$ stripe region around $n \approx 0.88$ and a $\eta = 0.1$ stripe region around $n \approx 0.78$, separated by a beat region consisting of wavevectors with both values of $\eta$. For low doping, $n \gtrsim 0.92$, the system is collinearly ordered, but not in a perfect stripe pattern. The charge and spin patterns in this regime resemble those displayed in panel (h) of Fig.~\ref{fig: Panel_With_Different_Orders}. As explained in Ref.~\cite{Zaanen1989}, this is likely due to the fact that longer wavelengths cannot be accommodated on a $20 \times 20$ lattice. Indeed, when repeating the calculations on a $48 \times 48$ system, we found a stripe ordered phase at a density of $n = 0.96$ for $T\leq 0.05$. The stripes we find in this regime do not appear in the shape of a simple cosine as in Eq.~\eqref{eq: PERFECT STRIPES}, but their profile is generally sharper. We find two almost half-filled regions separated by two hole-doped domain walls, which extend over 5-10 sites each. Since the domain walls are usually perfectly parallel to the $x$ or $y$ axis, and all spins are collinear, we still refer to this region as stripe ordered.
For $t^\prime = -0.3t$, there are two patches of stripe order, one centered around $n \approx 0.8$ corresponding to $\eta = 0.1$ and one centered around $n \approx 0.7$, with $\eta = 0.15$. Unlike in the $t^\prime = 0$ and $-0.15t$ case, there is a regime of collinear bidirectional stripes centered around filling $n = 0.6$, with $\eta = 0.2$,
followed by two regions of coplanar bidirectional stripes with $\eta = 0.25$ and $0.30$. These bidirectional stripe phases have a lower energy than the optimal spirals which we found from the momentum space calculation, but to confirm that they remain stable in the thermodynamic limit and are not artifacts of the finite system size, we repeated the calculation for four $(n,T)$ points in each of the three bidirectional stripe regimes for lattice sizes $28 \times 28$ and $40 \times 40$. In all cases, bidirectional stripe phases with a minimal free energy were found, so we believe that they will remain stable also in the thermodynamic limit.


\subsubsection{Magnetization}

\begin{figure}[t]
\centering
 \includegraphics[width=0.5\textwidth]{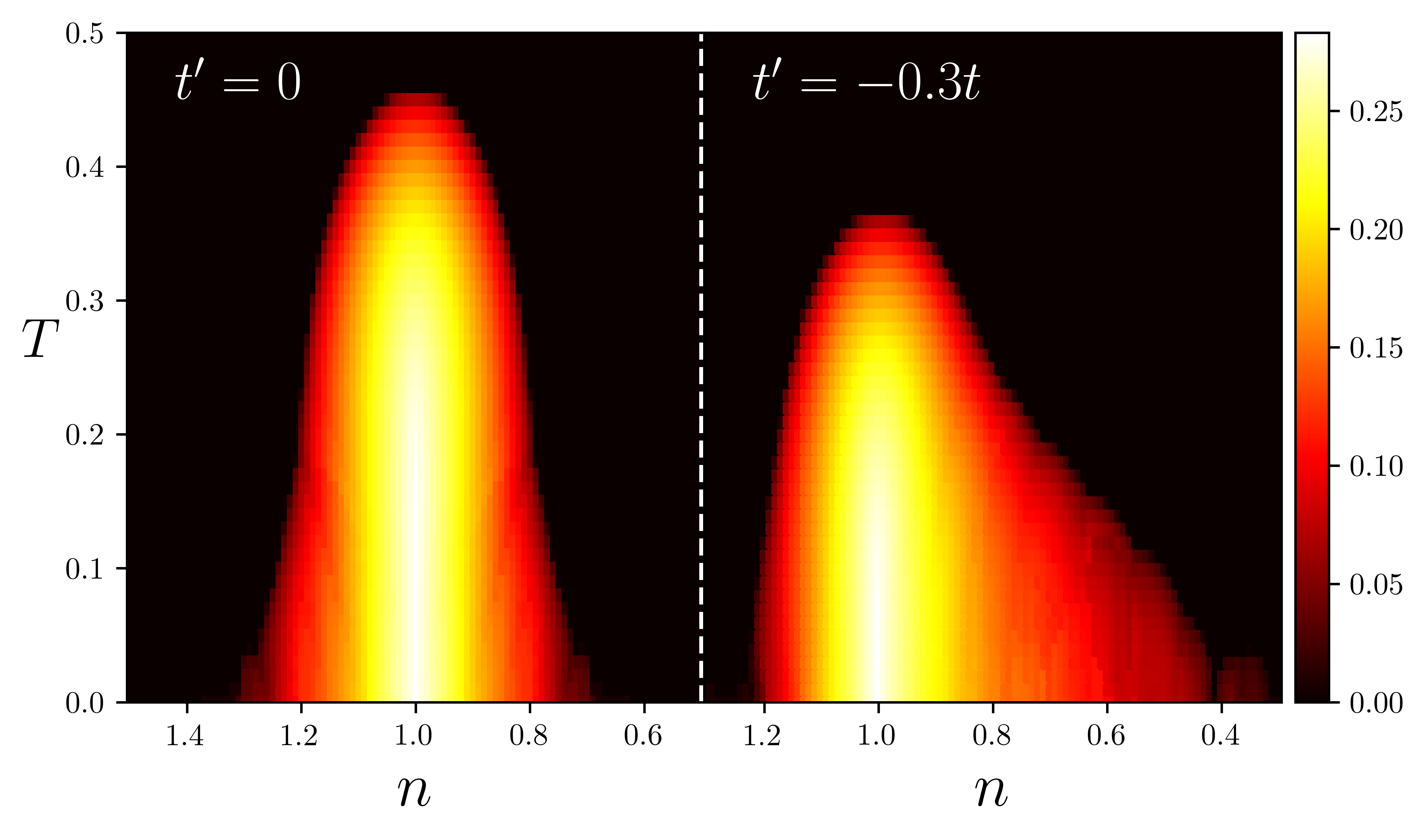}
 \caption{Average magnetization per site for $U = 3t$, $t'=0$ (left), and $t'=-0.3t$ (right) on a $20 \times 20$ lattice.}
\label{Magnetizations_Other_tprime}
\end{figure}
In Fig.~\ref{Magnetizations_Other_tprime}, we show the average magnetization per site for $t' = 0$ (left panel) and $t' = -0.3t$ (right panel). The magnetization is maximal near half-filling. The transition from the paramagnetic phase to the ordered phases seems again smooth, as in the $t^\prime = -0.15t$ case. However, due to discrete nature of our $(n,T)$ grid, we cannot exclude weak first order transitions.


\subsubsection{Energy considerations}

\begin{figure}[t]
\centering
 \includegraphics[width=0.5\textwidth]{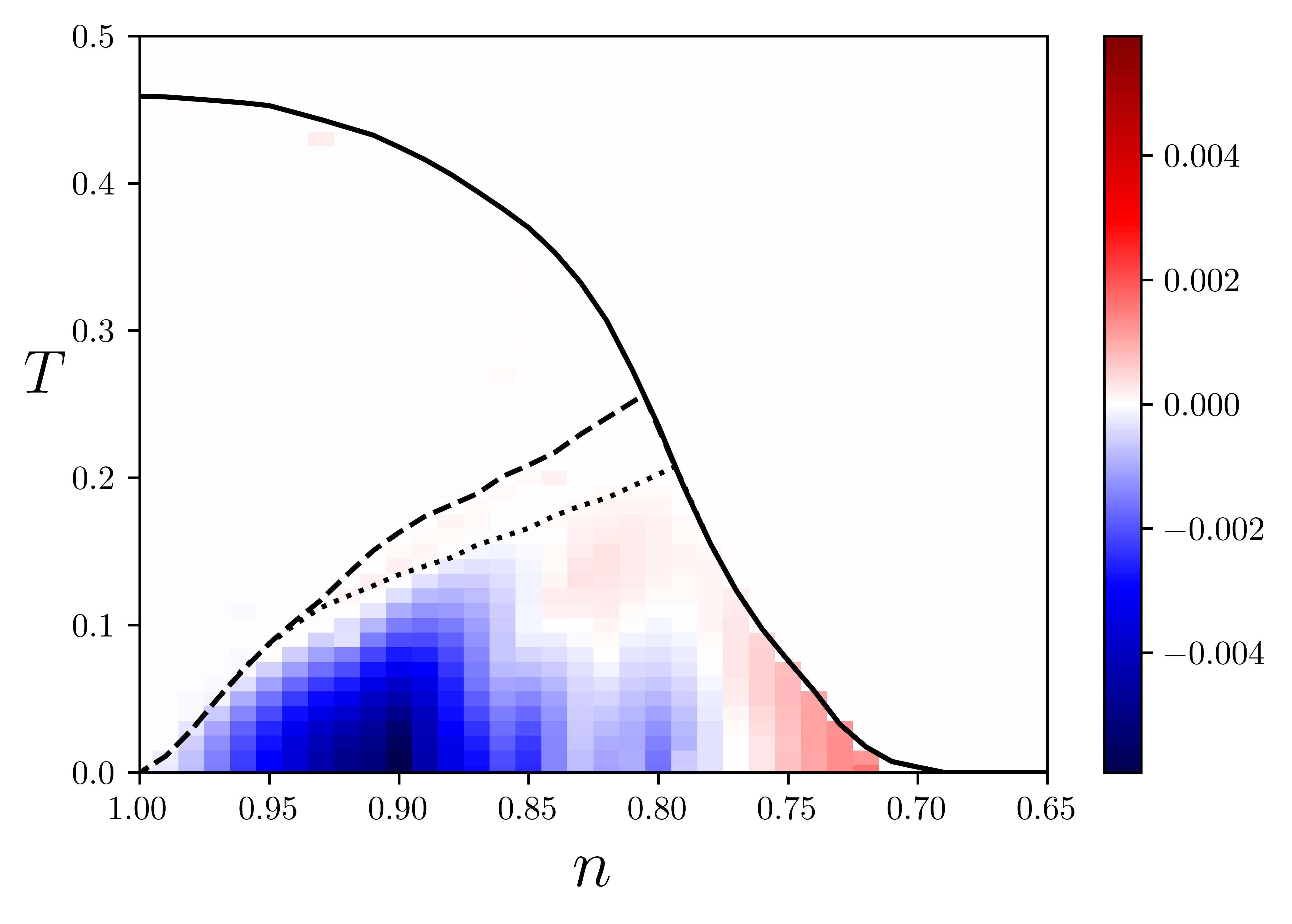}
 \caption{Energy difference $F_\mathrm{fs} - F_\mathrm{is}$ between the real space calculation on a $20 \times 20$ lattice and the momentum space calculation in the thermodynamic limit as in Fig.~\ref{EnergyDiagram015}, for $U = 3t$ and $t' = 0$.}
\label{EnergyDiagram0}
\end{figure}
\begin{figure}[t]
\centering
 \includegraphics[width=0.53\textwidth]{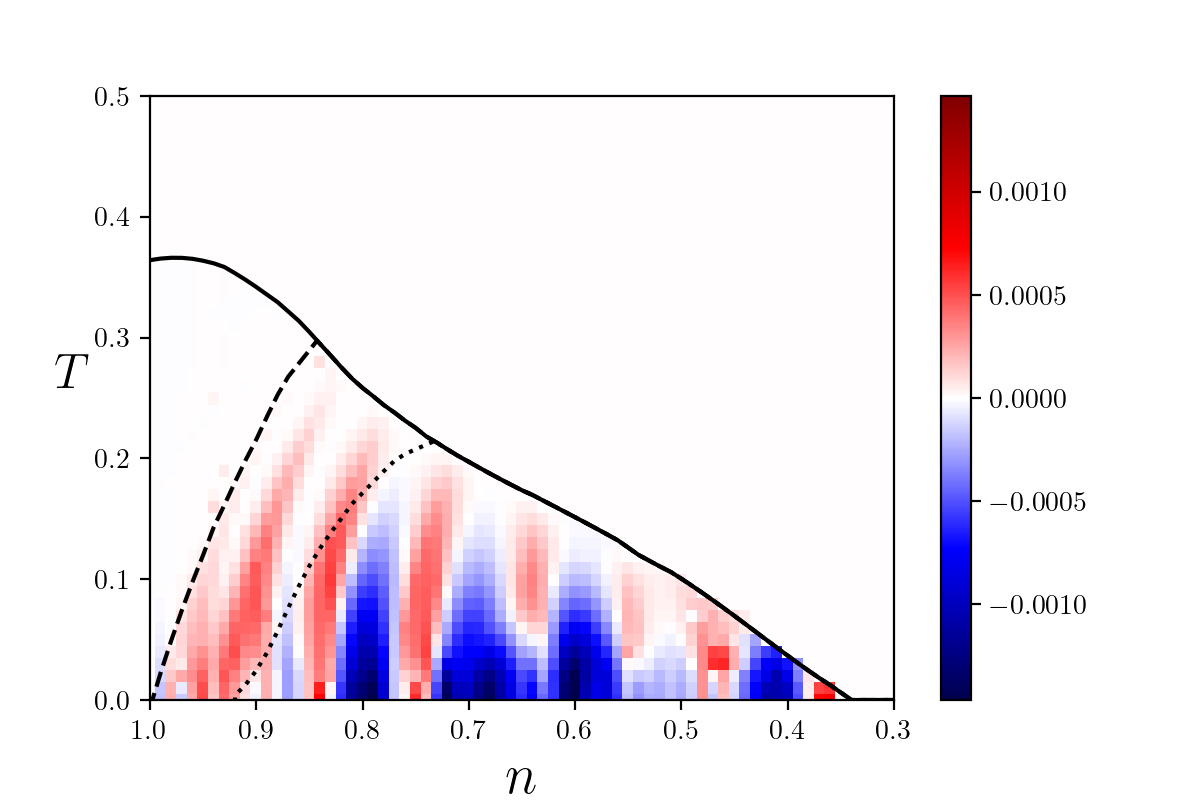}
 \caption{Energy difference $F_\mathrm{fs} - F_\mathrm{is}$ between the real space calculation on a $20 \times 20$ lattice and the momentum space calculation in the thermodynamic limit as in Fig.~\ref{EnergyDiagram015}, for $U = 3t$ and $t' = -0.3t$.}
\label{EnergyDiagram30}
\end{figure}
We have calculated the free energy differences between the real space calculation on a $20 \times 20$ lattice and the momentum space calculation in the thermodynamic limit, $F_\mathrm{fs} -F_\mathrm{is}$, in the magnetic regime for $t' = 0$ and $t' = -0.3t$, analogously to section~\ref{sec: EnergyConsiderations}. The results are shown in  Figs.~\ref{EnergyDiagram0} and \ref{EnergyDiagram30}, respectively.

In both plots, we see that the differences in free energies in the N\'eel and spiral regimes are almost zero. There is generally an energy gain in the stripe ordered phases and an energy loss in the beat and strange ordered phases, as for $t'=-0.15t$ in Sec.~\ref{sec: EnergyConsiderations}. For $t' = -0.3t$, we also see an energy gain in the collinear and coplanar bidirectional stripe phases at low fillings, compared to the spiral phase.

These observations reaffirm our expectation that the beat and strange states disappear from the phase diagram in the thermodynamic limit, while the N\'eel, spiral and stripe states remain stable.


\subsubsection{Density of states for $n=\frac{7}{8}$, $T = 0$ and $t' = 0$}

\begin{figure}[t]
\centering
 \includegraphics[width=0.5\textwidth]{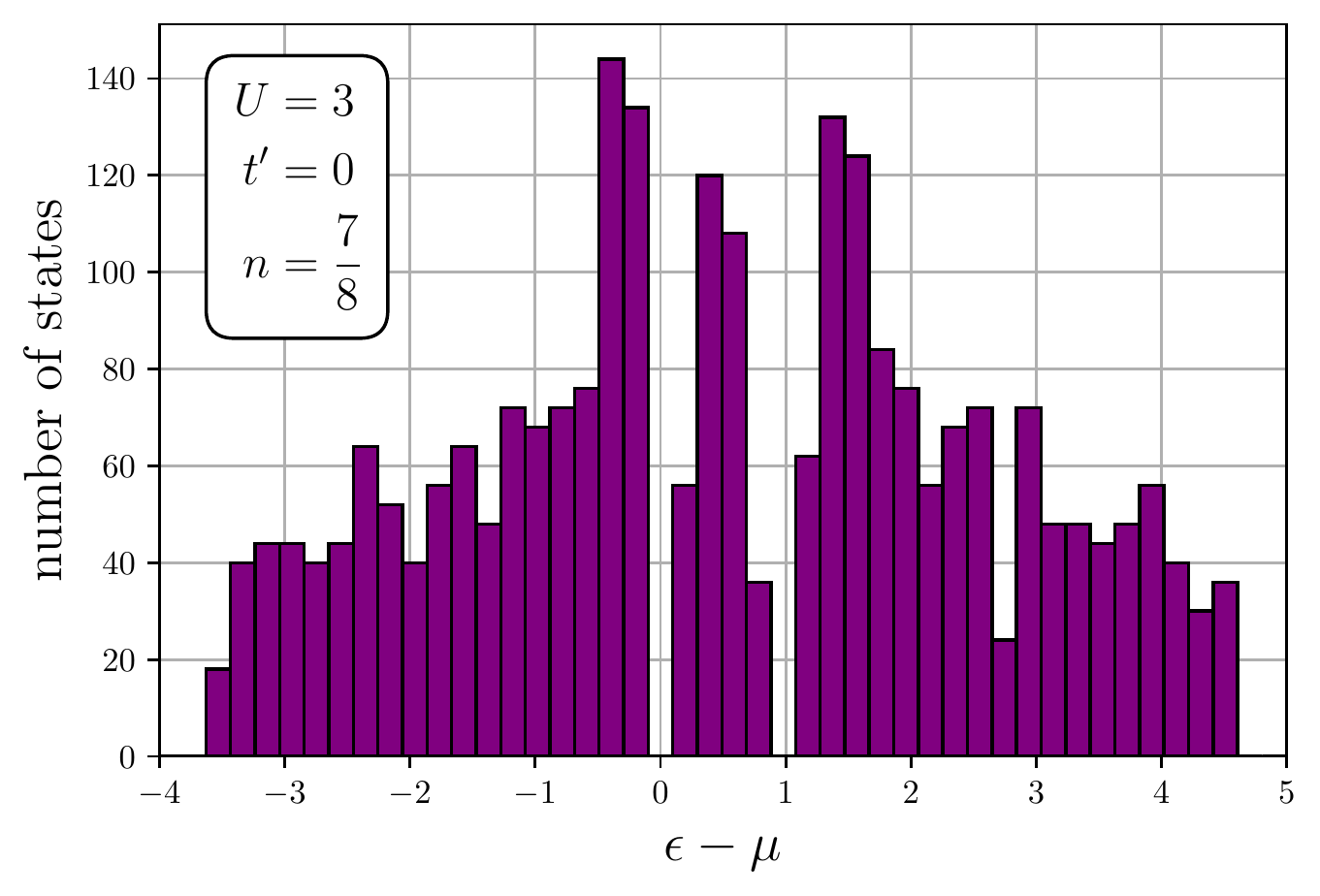}
 \caption{Density of states histogram for $n = \frac{7}{8}$, $T = 0$, $t' = 0$ and $U = 3$, calculated on a $40 \times 32$ lattice.
 The states are counted in equidistant bins in the energy range between -4 and 5.}
 \label{LDOS_eighthsdoping}
\end{figure}
A particular point in parameter space has received special attention in recent years: the ground state ($T=0$) at a density $n=\frac{7}{8}$ for the two-dimensional Hubbard model with pure nearest neighbor hopping. Exact numerical ground state techniques have provided fairly convincing evidence that for strong coupling the ground state is an insulating stripe state in this case \cite{Zheng2017,Qin2020}.
Hence, we investigate this special point in more detail, too, keeping however a moderate interaction strength $U = 3t$, in line with our Hartree-Fock approximation. To allow for a variety of distinct commensurabilities, and because stripes with wavelengths 8 and 16 were found by different numerical methods at $n = \frac{7}{8}$~\cite{Zheng2017, Hwang2018, Tocchio2019, Jiang2019, Qin2020},
we work on an anisotropic lattice with $\mathcal{N}_x = 40, \mathcal{N}_y = 32$ in this case.

The solution of the real-space mean-field equations unambiguously yieds unidirectional stripes with an incommensurability $\eta = \frac{1}{16}$ along the $y$-axis.
To see whether this stripe ground state is an insulator, we have computed the distribution of energy levels, which is a finite size proxy to the density of states.
In Fig.~\ref{LDOS_eighthsdoping}, we show a histogram of the distribution of the energy levels $\epsilon_l$, shifted by the chemical potential (placed half-way between the highest occupied and the lowest unoccupied state). The histogram is defined with equidistant bins between -4 and 5. We see that there are no energy levels close to the chemical potential, that is, there is a finite band gap with size $\Delta \approx 0.25t$.
The state we find for this parameter set is therefore indeed an insulator.
There are other insulating points in the ground state, but metallic behavior is much more frequent.


\subsection{Phase diagrams in the thermodynamic limit}

\begin{figure}[t]
\centering
 \includegraphics[width=0.5\textwidth]{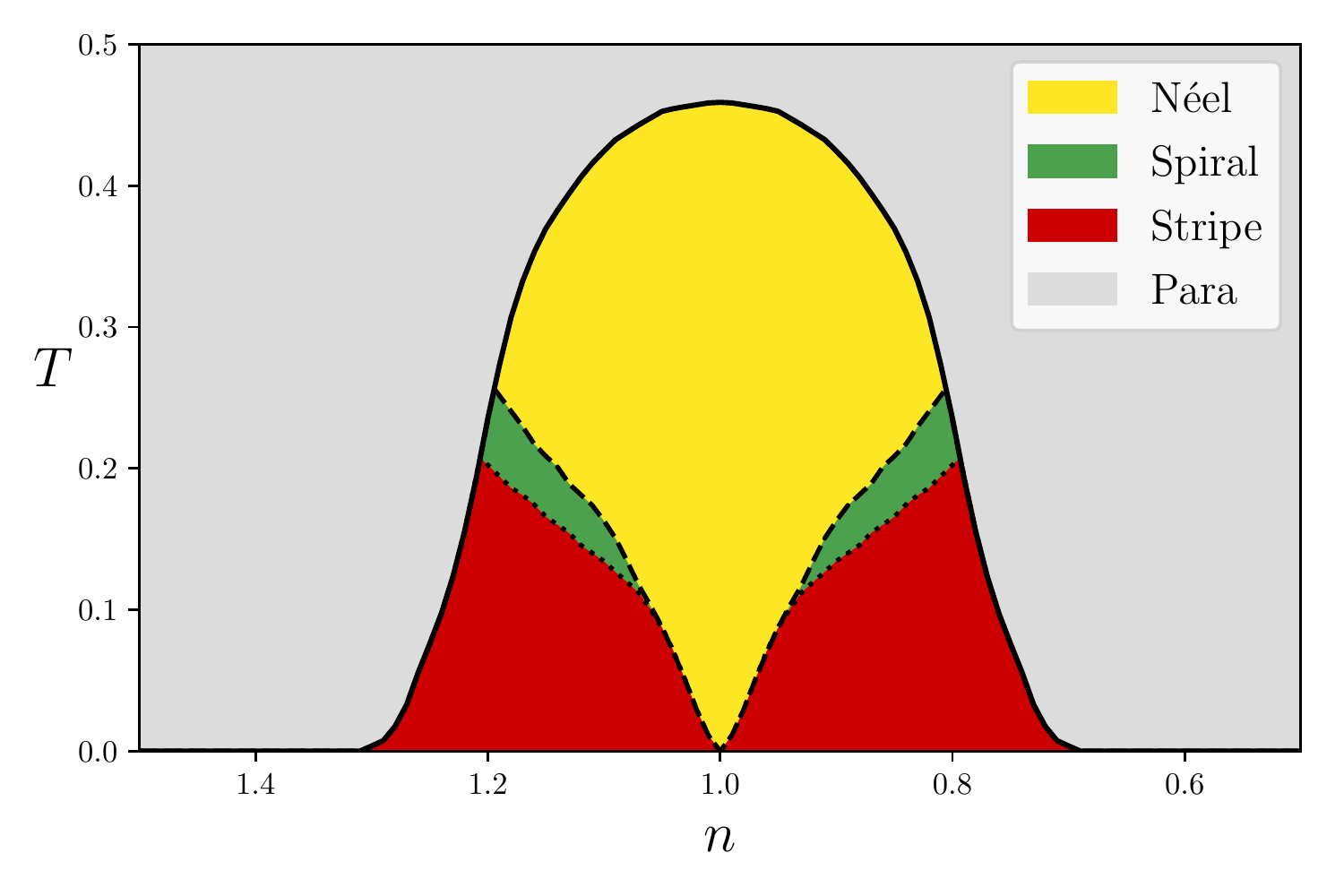}
 \caption{$(n,T)$ phase diagram for $t' = 0$ and $U = 3t$ in the thermodynamic limit with N\'eel, spiral, and unidirectional stripe order. Because of the Hamiltonian's particle-hole symmetry, this phase diagram is symmetric.}
\label{PhaseDiagram_tprime0_pretty}
\end{figure}

\begin{figure}[t]
\centering
 \includegraphics[width=0.5\textwidth]{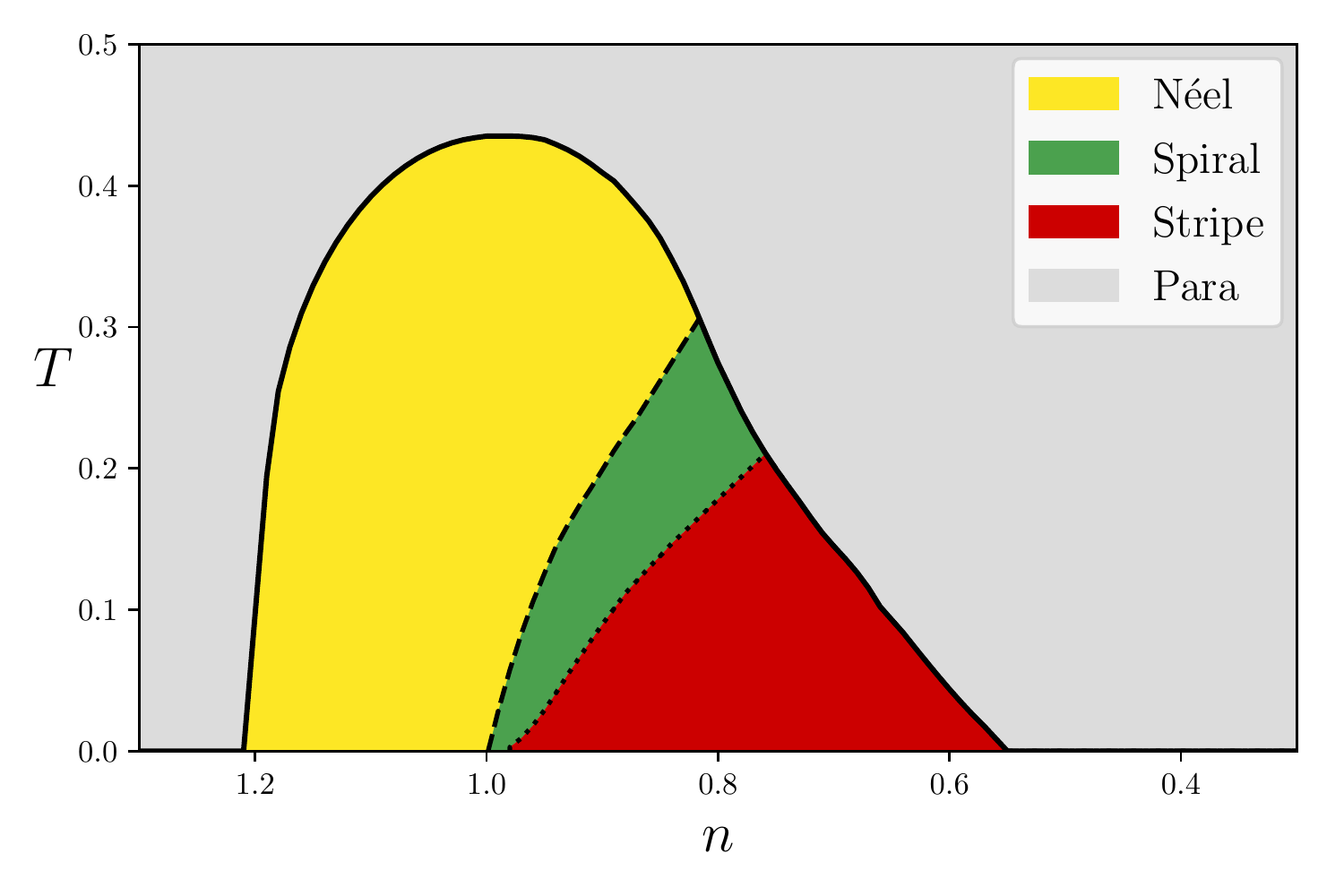}
 \caption{$(n,T)$ phase diagram for $t' = -0.15t$ and $U = 3t$ in the thermodynamic limit with N\'eel, spiral, and unidirectional stripe order. There is a narrow spiral regime at small hole doping even at $T=0$.}
\label{PhaseDiagram_tprime15_pretty}
\end{figure}

\begin{figure}[t]
\centering
 \includegraphics[width=0.5\textwidth]{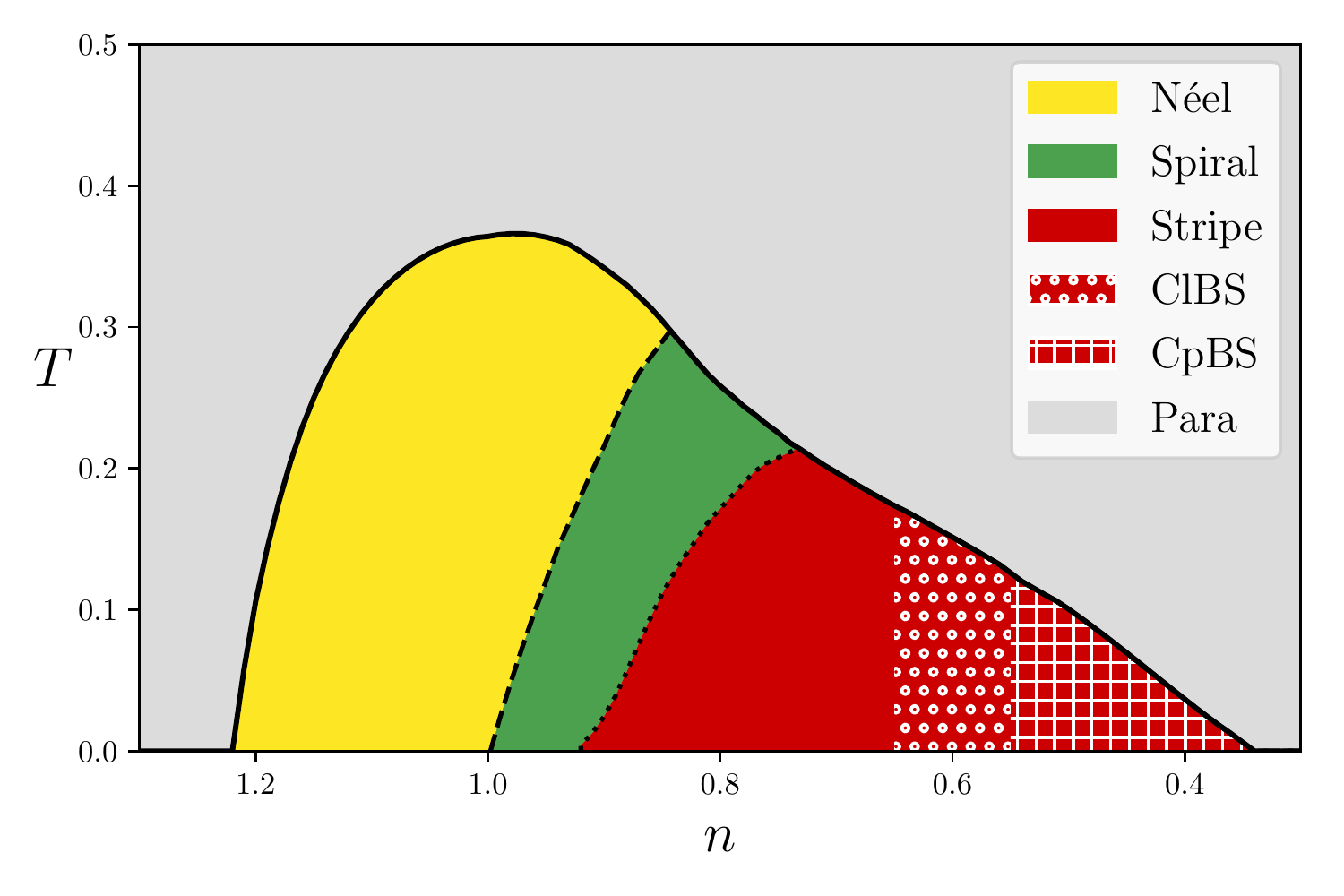}
 \caption{$(n,T)$ phase diagram for $t' = -0.3t$ and $U = 3t$ in the thermodynamic limit with N\'eel, spiral, and stripe order.
 Its topology is similar to the $t^\prime = -0.15$ case, but the spiral regime is larger, especially at low temperatures, and besides unidirectional stripes there are also regimes of collinear and coplanar bidirectional stripes.}
\label{PhaseDiagram_tprime30_pretty}
\end{figure}

Collecting the insights gained from the real space and momentum space solution of the mean-field equations discussed above, we now present our expectation for the mean-field phase diagrams of the two-dimensional Hubbard model in the thermodynamic limit.
These final $(n,T)$ phase diagrams for $U = 3t$ and $t' = 0$, $-0.15t$ and $-0.3t$ are shown in Figs.~\ref{PhaseDiagram_tprime0_pretty}, \ref{PhaseDiagram_tprime15_pretty} and \ref{PhaseDiagram_tprime30_pretty}, respectively.

Based on the energy considerations and the analysis of the dominant wavevectors in each phase, we concluded that neither beat- nor strange states will remain stable in the thermodynamic limit. There is no indication that the N\'eel and spiral states are unstable within the regions shown in the phase diagrams. The divergence of the charge susceptibility at the low temperature boundary of the spiral regime (dotted line) indicates a transition to a stripe state, which is indeed found within the real space calculation whenever the optimal wave vector can be accommodated on the finite lattice.

At zero temperature, the N\'eel state is stable for densities $n \geq 1$ for $t' \leq -0.15t$, and a spiral phase extends from half-filling to a hole doping $p = 0.02$ for $t' = -0.15t$, and to $p = 0.09$ for $t' = -0.3t$, while for larger hole doping stripe order prevails. For $t' = 0$ the N\'eel ground state is unstable toward stripe order at any finite doping.
For $t' = -0.3t$ we also found collinear bidirectional stripes and coplanar bidirectional stripes on the $20 \times 20$ lattice, which remain stable also for larger systems. It is difficult to determine the exact domain borders of the three stripe variants in the thermodynamic limit. On the $20 \times 20$ lattice, the system exhibits ClBS order in a doping range of $p \approx 0.35$ to $0.45$ and CpBS order in a doping range of $p \approx 0.45$ to $0.65$.


\section{Conclusion} \label{sec: Conclusion}

We have performed an unbiased mean-field analysis of magnetic and charge orders in the two-dimensional Hubbard model, both at zero and finite temperature. Fully unrestricted Hartree-Fock calculations on large finite lattices (from $20 \times 20$ to $48 \times 48$) have been complemented by momentum space solutions restricted to N\'eel and circular spiral states in the thermodynamic limit.
The Hubbard interaction was fixed at the moderate value $U = 3t$, while three distinct ratios of next-nearest neighbor to nearest neighbor hopping amplitudes were considered: $t'/t = 0$, $-0.15$, and $-0.3$.
The magnetic states were classified by a systematic scheme based on the dominant Fourier components $\Vec{\mathcal{S}}_\bq$ of the spin texture $\langle \Vec{S}_j \rangle$.
All but N\'eel and circular spiral states entail charge, in addition to spin, order.

On the finite lattices a whole zoo of magnetic states has been found, some of which looking quite messy. However, comparing solutions on different lattices, analysing the condensation energies, and comparing to N\'eel and spiral solutions in the thermodynamic limit, we could show that the more complex states are finite size artifacts, which are related to the limited choice of ordering wave vectors on finite lattices. On a finite lattice with periodic boundary conditions the optimal wave length of a spin (and charge) density wave can be accommodated only if the lattice size is an integer multiple of that wave length. If the mismatch is too big, the magnetic and charge order assume complex structures, for example with charge stripes closing to a ring \cite{Zaanen1989}. These ``compromise'' states disappear as the lattice size increases.

In the thermodynamic limit, only N\'eel, circular spiral, and stripe states are present. The latter are usually unidirectional, but can be bidirectional for $t'=-0.3t$. All phases are uniform, that is, we never find phase separation.
The final $(n,T)$ phase diagrams for $t'=0$, $-0.15t$, and $-0.3t$ are shown in Figs.~\ref{PhaseDiagram_tprime0_pretty}, \ref{PhaseDiagram_tprime15_pretty}, and \ref{PhaseDiagram_tprime30_pretty}, respectively.
The boundaries of the N\'eel and spiral regimes can be determined with high accuracy directly in the thermodynamic limit from the momentum space solution of the mean-field equations restricted to these phases, complemented by an analysis of the RPA charge and spin susceptibilities. The susceptibilities are positive for all wave vectors inside the N\'eel and spiral domains, and divergences occur exclusively due to the Goldstone modes \cite{Bonetti2022, Bonetti2022ward}.
The charge distribution is uniform in the entire N\'eel and spiral regime.
The transition to the stripe regime is marked by a divergence of the charge susceptibility.

In the ground state, N\'eel order is limited to half filling if $t'=0$, and extends to the electron doped regime ($n > 1$) for $t' = -0.15t$ and $t' = -0.3t$. N\'eel order is replaced by spiral or stripe order for arbitrarily small hole doping at $T=0$. Note, however, that N\'eel order can be stable for low hole doping and negative $t'$ for weaker bare or renormalized interactions \cite{Chubukov1992, Chubukov1995, Yamase2016}.
In most parts of the phase diagram, the spatial dependence of the spin and charge patterns is well described by simple trigonometric functions (sine and cosine) with few wave vectors. Only for $t'=0$ and low doping a sharp real space profile with antiferromagnetic domains separated by hole-rich domain walls develops, corresponding to a sizable contribution from higher harmonics in a Fourier decomposition.
While doped N\'eel and spiral states are always metallic, we find some insulating spots in the stripe regime, in particular for the intensively studied special case $t' = 0$ and $n = \frac{7}{8}$, in qualitative agreement with exact numerical calculations on finite lattices \cite{Zheng2017, Hwang2018, Tocchio2019, Jiang2019, Qin2020}.

A static mean-field calculation is clearly not applicable to the physics of cuprate superconductors, which should be described by the Hubbard model (or extensions) at strong coupling. Nevertheless, some broad features we obtain are in line with experimental observations in cuprates \cite{Keimer2015}: N\'eel order at and near half-filling, with a broader N\'eel regime for electron doped cuprates, incommensurate magnetic correlations with wave vectors of the form $\bQ = (\pi - 2\pi\eta,\pi)$ for hole doped cuprates, and charge and/or stripe order for some of them.

The mean-field analysis presented in our paper could be extended in various directions. At moderate coupling, substantial quantitative improvement could be achieved and the interplay with $d$-wave pairing could be studied by using a renormalized mean-field theory where effective couplings obtained from a functional renormalization group flow are used instead of the bare Hubbard interaction \cite{Wang2014, Yamase2016, Vilardi2020}.
An extension of dynamical mean-field theory (DMFT) to inhomogeneous systems, the iDMFT \cite{Potthoff1999}, can capture arbitrary spin and charge order patterns at strong coupling. Inhomogeneous DMFT studies of spin and charge order in the two-dimensional Hubbard model have already been performed \cite{Fleck2000, Raczkowski2010, Peters2014}, but so far only at zero temperature and for pure nearest neighbor hopping.
Finally one could take spin fluctuations into account to replace the magnetic state with long-range order by a fluctuating magnet with short-range (and possibly topological) order by extending SU(2) gauge theories with N\'eel and spiral ordered chargons \cite{Scheurer2018, Sachdev2019review, Bonetti2022gauge} to stripes.


\section*{Acknowledgments}

We are grateful to Andrey Chubukov and Hiroyuki Yamase for useful discussions.


\begin{appendix}

\section{Computational details} \label{sec: ImplementationDetails}

We provide here additional information about the procedure we used to obtain converged states for a given set of parameters $U, t^\prime, T, n ,\mathcal{N}_x, \mathcal{N}_y$.
The code we developed to perform the calculations is publicly available~\cite{codeLink}.


\subsection{Mixing}

To improve the convergence of the iterative scheme, we employ a linear mixing between each iteration. Instead of constructing the new Hamiltonian with the expectation values as obtained from Eq.~\eqref{eq: exp value T=0} or~\eqref{eq: exp value T!=0}, we instead use 60\% of the new 
expectation values and keep 40\% of the current ones. The mixing has proven to be essential to ensure convergence and avoid getting trapped in local minima.


\subsection{Convergence criteria}

We consider a state to be converged, if either all the parameters $\Delta_{j\alpha}$ from Eq.~\eqref{eq: GapVectorDefiningEquation} change by less than $10^{-8}$ between two iterations, or after 3500 iterations. 
Reducing the threshold value below $10^{-8}$ does not affect the results.
The second convergence criterion was chosen to ensure that the computation will eventually terminate. 3500 iterations is usually more than enough to obtain a state which is already clearly defined and can be classified as the same type of magnetic order as the fully converged state. Since we additionally calculated each $(n,T)$ pixel multiple times with various random initial conditions, we are confident to have classified the true global minimum for each parameter set.


\subsection{Scaling of the code with $\mathcal{N}$}
In each iteration the step which takes by far most of the time is the diagonalization of the quadratic Hamiltonian, Eq.~\eqref{MeanFieldHamburger}.
This matrix diagonalization scales roughly as $\mathcal{N}^3$. The matrix is very sparse, so there are more efficient algorithms for obtaining the lowest eigenstates.
However, we make use of \textit{all} the eigenvalues and eigenvectors, and most of the commonly used eigensolvers for sparse matrices are only efficient to obtain few of the lowest energy eigenstates.


\section{Classification of states}  \label{app: class of states}

We here give an overview of how we classify states on the $20 \times 20$ lattice via the Fourier components $\Vec{\mathcal{S}}_{\bq}$ introduced in Eq.~\eqref{Classification_state_ala_Sachdev}. The classification is usually not too sensitive to the chosen thresholds and the used norms.

If we find, that 
\begin{equation}
    \frac{1}{\mathcal{N}}\sqrt{\sum_{\bq} |\Vec{\mathcal{S}}_{\bq}|^2} < 2.5\times 10^{-7}, 
\end{equation}
we classify the state as paramagnetic.

To classify the other states, we first note that
the $\Vec{\mathcal{S}}_{\bq}$ with the largest norms usually correspond to wave vectors of the form
\begin{equation}\label{eq: Form of Fourier modes}
\begin{split}
    \bq_x &= (\pi-2\pi\eta_x,\pi),\\
    \bq_y &= (\pi,\pi-2\pi\eta_y),
\end{split}
\end{equation}
with variable $\eta_{x,y}$. Note that we use the notation $\bq_x$ ($\bq_y$) to indicate the \textit{class} of vectors of the form~\eqref{eq: Form of Fourier modes}, while $\bQ_x$ ($\bQ_y$) denotes a \textit{single} wavevector with a fixed value of $\eta_x$ ($\eta_y$). We also checked for diagonal order of the form $\bq_{xy} \equiv(\pi-2\pi\eta,\pi-2\pi\eta)$, but we found almost no states with these modes as dominant contributions (less than 40 $(n,T)$ pixels in all three phase diagrams) and when we found them, the order did not remain stable when repeating the calculations on a $40\times40$ lattice.

To determine the $\ell$ dominant $\bq$-modes, labeled as $\bq_i$, of a state, we used the condition
\begin{equation}
 \frac{\sqrt{\sum_{i=1 }^l  |\Vec{\mathcal{S}}_{\bq_i}|^2 + |\Vec{\mathcal{S}}_{-\bq_i}|^2} }{\sqrt{\sum_{\bq \in \{\bq_x, \bq_y \} }  |\Vec{\mathcal{S}}_{\bq}|^2} } > 95\% ,
\end{equation}
where $\bq_1 \hdots \bq_l$ are the subset of wavevectors $\{\bq_x, \bq_y \}$ with the largest amplitudes $|\Vec{\mathcal{S}}_{\bq_i}|$ in descending order, and we always considered the $\bq$ and $-\bq$ modes as a single mode, since $\vec{\mathcal{S}}_{-\bq} = \vec{\mathcal{S}}^*_{\bq}$.

For most states we find that only one or two modes have a significant Fourier component. If three or more modes contribute, we label the state as displaying either strange or other collinear order.
To check whether a state is collinearly ordered, we first rotate all  $\Vec{\mathcal{S}}_{\bq}$ so that the real part of  $\Vec{\mathcal{S}}_{\bq}$ with the biggest norm points in $x$ direction. We then say that a state is collinear, if
\begin{equation}
\frac{1}{\mathcal{N}} \sqrt{\sum_{\bq}  |\mathcal{S}^y_{\bq}|^2} < 2.5 \times 10^{-7},
\end{equation}
and the same for $\mathcal{S}^z_{\bq}$.
We proceed analogously, if the $\bq$-modes of the $\bq_x$ or $\bq_y$ form make up less than 70 $\%$ of the total norm of the Fourier transformation, that is if
\begin{equation}
\frac{\sqrt{\sum_{\bq \in \{\bq_x, \bq_y \} }  |\Vec{\mathcal{S}}_{\bq}|^2} }{\sqrt{\sum_{\bq}  |\Vec{\mathcal{S}}_{\bq}|^2} }< 70\%.
\end{equation}

If we find only two contributing modes $\bQ_1,\bQ_2 \in \{\bq_x,\bq_y \}$,
we proceed as follows:

If $\bQ_1$ and $\bQ_2$ are both in $\{\bq_x\}$ or both in $\{\bq_y\}$, we consider the state a beat state. Typically, $\bQ_1$ and $\bQ_2$ correspond to adjacent allowed values of $\eta$.

If $\bQ_1$ and $\bQ_2$ have the form $\bQ_1 = (\pi-2\pi\eta,\pi)$ and $\bQ_2 = (\pi,\pi-2\pi\eta)$, we rotate the vectors so that the real part of $\Vec{\mathcal{S}}_{\bQ_1}$ points in $x$ direction, then divide each vector by the norm of $\Vec{\mathcal{S}}_{\bQ_1}$. We then check how the $\Vec{\mathcal{S}}_{\bQ_1}$ are oriented relative to each other with an error tolerance of $5 \cdot 10^{-3}$.
For example, for coplanar bidirectional stripes (see Sec.~\ref{ClassificationSection}), we check whether $\Vec{\mathcal{S}}_{\bQ_1} \cdot \Vec{\mathcal{S}}_{\bQ_2} = 0$ and $\mathcal{S}^y_{\bQ_1} = \mathcal{S}^z_{\bQ_1} = 0$ and $|\Vec{\mathcal{S}}_{\bQ_1}| = |\Vec{\mathcal{S}}_{\bQ_2}|$, each equality up to a tolerance of 0.5\%.
We proceed analogously for collinear bidirectional stripes. If we find different orientations of $\Vec{\mathcal{S}}_{\bQ_1}$, $\Vec{\mathcal{S}}_{\bQ_2}$, as proposed in \cite{Sachdev2019}, we would classify the state as strange ordered, but this happens very rarely.

If we find only one contributing mode $\bQ$, we proceed analogously to identify stripes or spirals. If additionally $\eta = 0$, we classify the state as a N\'eel antiferromagnet.

\end{appendix}

\bibliography{main.bib}

\end{document}